\documentclass[twocolumn,amsmath,amssymb,floatfix,prb,showpacs,footinbib,superscriptaddress]{revtex4}
\usepackage{amsmath,amssymb,natbib,bm,graphicx,url,psfrag,times,eucal,enumitem}
\usepackage{color}

\def\OSUM{{%
    \setbox0\hbox{$\displaystyle\sum$}%
    \rlap{\hbox to \wd0{\hss$\!\circlearrowleft$\hss}}\box0
}}

\def\otsum{\mathrel{%
    \mathchoice{\OTSUM}{\OTSUM}{\scriptsize\OTSUM}{\tiny\OTSUM}%
}}
\def\OTSUM{{%
    \setbox0\hbox{$\textstyle\sum$}%
    \rlap{\hbox to \wd0{\hss$\!\circlearrowleft$\hss}}\box0
}}

\newcommand{\NN}{\mathbb{N}}

\newcommand{\abs}[1]{|#1|}
\newcommand{\bra}[1]{\langle \, #1 \,|}
\newcommand{\ket}[1]{|\, #1 \, \rangle}
\newcommand{\bket}[2]{\langle \, #1 \,|\, #2 \, \rangle}
\newcommand{\boket}[3]{\langle\, #1 \,|\, #2 \,|\, #3 \,\rangle}

\newcommand{\be}{\begin{equation}}
\newcommand{\ee}{\end{equation}}
\newcommand{\bc}{\begin{center}}
\newcommand{\ec}{\end{center}}

\begin{document}
\title{Nematic quantum liquid crystals of bosons in frustrated lattices }
\author{Guanyu Zhu}
\affiliation{Departments of Physics and Astronomy, Northwestern University, Evanston, Illinois 60208, USA}
\author{Jens Koch}
\affiliation{Departments of Physics and Astronomy, Northwestern University, Evanston, Illinois 60208, USA}
\author{Ivar Martin}
\affiliation{Materials Science Division, Argonne National Laboratory, Argonne, Illinois 60439, USA}
\date{\today}

\begin{abstract}
The problem of interacting bosons in frustrated lattices is an intricate one due to the absence of a unique
minimum in the single-particle dispersion where macroscopic number of bosons can condense. Here we
consider a family of tight-binding models with macroscopically degenerate lowest energy band, separated from
other bands by a gap. We predict the formation of exotic states that spontaneously break rotational symmetry at
relatively low filling. These states belong to three nematic phases:  Wigner crystal, supersolid, and superfluid. The Wigner crystal phase
is established exactly at low filling. Supersolid and superfluid phases, at larger filling, are obtained by making
use of a projection onto the flat band, construction of an appropriate Wannier basis, and subsequent mean-field
treatment. The nematic superfluid that we predict is uniform in real space but has an anisotropic momentum
distribution, providing a novel scenario for Bose condensation with an additional nematic order. Our findings
open up a promising direction of studying microscopic quantum liquid crystalline phases of bosons.

\end{abstract}
\maketitle

\section{Introduction}

Interactions take the center stage when the kinetic energy of particles is quenched. The most prominent example is the fractional quantum Hall effect, where Coulomb interaction lifts the degeneracy of a partially filled Landau level and leads to formation of topological fractionalized states. There, kinetic energy of particles  is quenched due to the magnetic field that localizes them within the magnetic length, leading to formation of flat bands -- the Landau levels.
Flat bands also appear in some tight-binding lattice models, as is the case for the kagome lattice. Hence, a question naturally arises -- what kind of exotic states can emerge in such systems?

The problem of interacting bosons in frustrated lattices with a flat band has been a subject of significant theoretical interest \cite{Huber:2006bc, Wu:2007iz, Bergman:2008ce,     Wang:2011cy, You:2012cr, Wang:2012gj,Sedrakyan:2014cs}.   For exactly degenerate single particle spectrum, there is no preferred momentum state for the Bose condensation to occur.   Depending on the specific model and parameter regime, bosonic ground states in such frustrated lattices may include chiral composite-fermion states of hard-core bosons \cite{Sedrakyan:2014cs, Yang:1993js, Kumar:2014fo} and chiral superfluid/Mott insulator states \cite{Zaletel:2014de} which spontaneously break time-reversal symmetry, fractional Chern insulators \cite{Wang:2011cy, Wang:2012gj}, and other exotic broken symmetry states \cite{Huber:2006bc, Bergman:2008ce,  Wu:2007iz,  Huse:1992vl, You:2012cr, Moller:2010ce}.

From the experimental standpoint, rapid advance of artificial condensed matter systems such as cold atoms and interacting photons in circuit QED system have enabled not only the realization of geometrically frustrated lattices but also lattices subject to synthetic gauge fields \cite{jaksch_creation_2003-1, Dalibard:2011gg,  0034-4885-77-12-126401, Gerbier:2010ho, Lin:2012hn,  Aidelsburger:2013ew, Ji:2014jh, Jotzu:2015kz, koch_time-reversal_2010, Nunnenkamp:2011dm, Hafezi:2013vp, Hafezi:2012ui}.   These recent developments make the search for accurate theoretical approaches and concrete proposals timely.

In this paper,  we focus on a \emph{modified} kagome lattice model, constructed in a way to allow controlled treatment thanks to a spectral gap.  This gap can be generated by inserting an additional gauge flux into each hexagon of the kagome lattice \cite{Katsura:2010di}. We show that the single-particle eigenstates comprising the gapped flat band can be chosen as localized \textit{loop states}, which typically break the lattice $D_6$ point group symmetry.  Previous works in the context of the Bose-Hubbard model have primarily considered the \emph{simple} kagome lattice \cite{Sachdev:1992kp, Yang:1993js,  Schulenburg:2002bv, Huber:2006bc, Huse:1992vl, balents_fractionalization_2002, You:2012cr, Capponi:2013cc, Punk:2014be, Kumar:2014fo}. There,  the lowest band is flat but gapless since it is in contact with another band at the $\Gamma$ point \cite{Bergman:2008ce, Huber:2006bc, You:2012cr}. This makes the analysis of the interacting problem quite subtle due to the ability of particles to leak easily into the higher band.

The introduced gap enables a well-controlled projection onto the flat-band subspace \cite{Huber:2006bc, Zhitomirsky:2004ip,  Zhitomirsky:843638, PhysRevB.88.220510, Takayoshi:2013ws} in the weak-interaction regime, $U\ll\abs{t}$, and yields an effective low-energy Hamiltonian applicable to a wide filling range. This is analogous to the lowest-Landau level projection employed in the fractional quantum Hall effect.
Depending on the filling fraction of the lattice, we find three types of exotic nematic phases. At close packing of maximally compact loop states, a nematic Wigner crystal is the exact ground state of the system. In the specific case of $\pi$-flux and higher filling fraction, our mean-field treatment predicts transitions to a non-uniform nematic supersolid followed by a uniform nematic superfluid phase.

The nematic superfluid phase is quite unusual since it is not featureless but contains internal structure in its microscopic many-body wavefunction. The lattice rotational
symmetry is spontaneously broken due to the anisotropic correlations among the loop orbitals. Such anisotropic internal structure is encoded in the momentum distribution, i.e.~the
Fourier transform of the real-space correlation function. In addition to the standard delta-function peak, there is an anisotropic and squeezed continuous background in the
momentum distribution.  It clearly reveals a
novel nematic Bose condensation and can be detected through time-of-flight imaging in the context of ultra-cold atoms.  In addition, we show that the nematicity can also manifest itself in macroscopic quantities, namely the anisotropic superfluid stiffness tensor and superflow, which can be probed with phase imprinting
techniques in ultra-cold atom setups \cite{Isoshima:2000vh}.

From a broader perspective, the possibility of such microscopic liquid crystalline phases has been pointed out previously in the context of strongly correlated electronic materials \cite{Kivelson:1998cq, Fernandes:2014pt}. While in this article, we focus on translationally invariant/periodic states,  additional rich physics is associated with topological defects \cite{DeGennesLiquidCrystalBook} and warrants future studies.  

The paper is organized as follows. Section \ref{Model} shows the interacting boson model we study throughout the paper. Section \ref{single} discusses the single-particle eigenstates of the model, including the gapped flat band structure and the presence of localized loop eigenstates in real space and their specific properties such as ``flux quantization", with the detailed derivation shown in Appendix A.  In Section \ref{exactsolution}, we work out the exact Wigner-crystal ground states below the close-packing filling of loop states. In Section \ref{nematic}, we study the quantum phases beyond the close-packing filling. We apply a flat-band projection based on the construction of mutually orthogonal and spatially compact Wannier states, with the details of construction shown in Appendix C. We then perform a subsequent mean-field analysis in the Wannier basis, which predicts the existence of nematic superfluid and supersolid phases. In Section \ref{signature}, we show the novel signatures of the nematic superfluidity, namely the anisotropic momentum distribution and the anisotropic superflow. The detailed calculations of the momentum distribution can be found in Appendix F.  We conclude our work and provide a brief outlook in Section \ref{conc}.

\begin{figure}
\includegraphics[width=0.9\columnwidth]{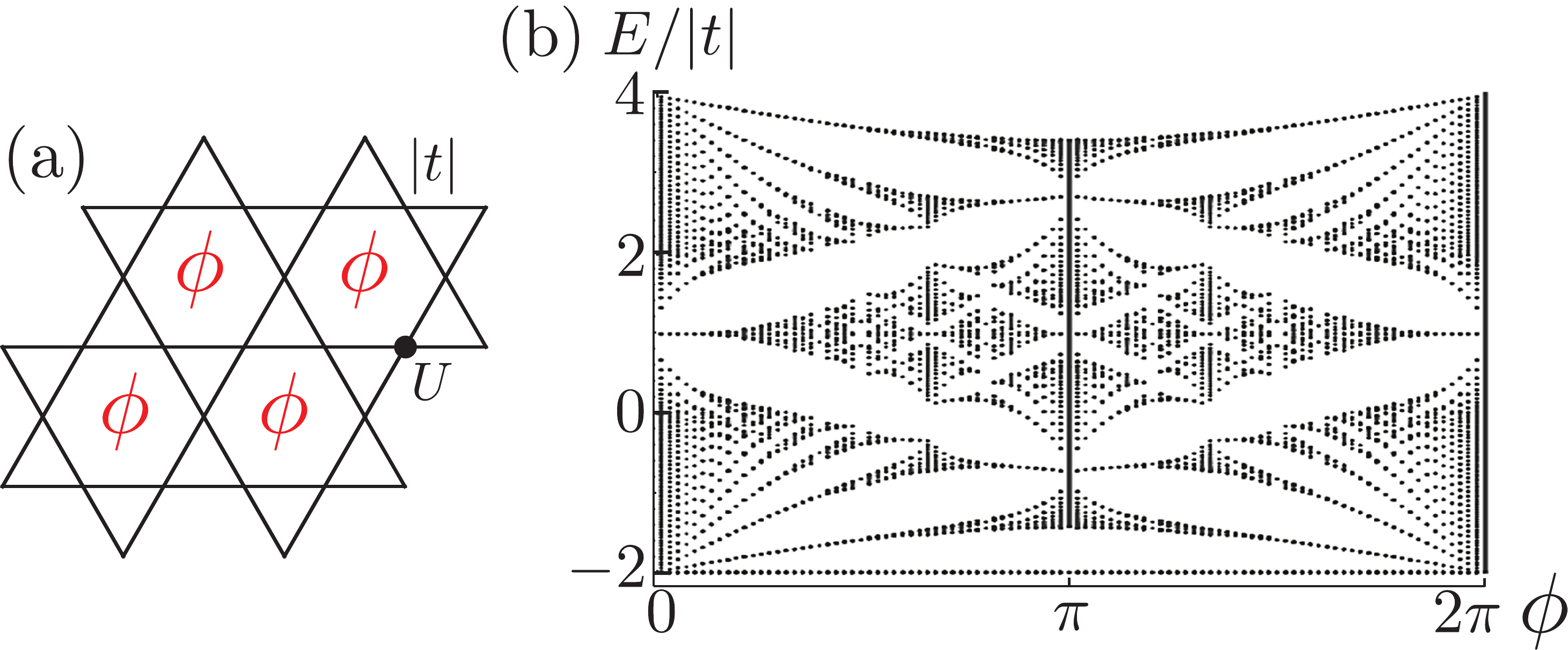}
\caption{(color online). (a) Kagome lattice with positive hopping and flux $\phi$ penetrating each hexagon.   (b)~Tight-binding energies as a function of $\phi$ (Hofstadter butterfly).   The lowest flat band is gapped for non-integer $\phi/2\pi$ and reaches its maximum gap size at $\pi$ flux.}  
\label{butterfly}
\end{figure}

\section{Model}\label{Model}
The Bose-Hubbard model on the kagome lattice subject to gauge flux [Fig.\ \ref{butterfly}] is described by the Hamiltonian
\be\label{Hamiltonian}
H= \sum_{\langle \mathbf{r},\mathbf{r'} \rangle} \left(\abs{t} e^{i A_{\mathbf{r} \mathbf{r}'}} b^\dag_{\mathbf{r'}} b_{\mathbf{r}} + \text{h.c.} \right) + U \sum_{\mathbf{r}} b^\dag_{\mathbf{r}} b^\dag_{\mathbf{r}} b_{\mathbf{r}} b_{\mathbf{r}},
\ee
where $b_{\mathbf{r}}^\dag$ creates a single boson on the site labeled $\mathbf{r}$. The first term is the tight-binding Hamiltonian $H_\text{tb}$ determining the band structure of the non-interacting bosons.  We denote the hopping amplitude by $\abs{t}$ to stress that it is positive (frustrated) \footnote{In ultracold-atom experiments, the natural negative hopping can be turned positive by threading triangles of the kagome lattice with a $\pi$ flux.}. The gauge potential $A_{\mathbf{r}\mathbf{r}'}$ is defined on each nearest-neighbor bond $\langle \mathbf{r}, \mathbf{r}' \rangle$ determines the flux 
$\phi=\otsum_{\langle \mathbf{r},\mathbf{r}' \rangle}A_{\mathbf{r}\mathbf{r}'}$ threading each plaquette in the lattice. The second term in $H$ captures the repulsive Hubbard interaction on each site with strength $U>0$.

\section{Single-particle eigenstates}\label{single}

\subsection{Gapped flat band in the presence of flux}

\begin{figure}
\includegraphics[width=1 \columnwidth]{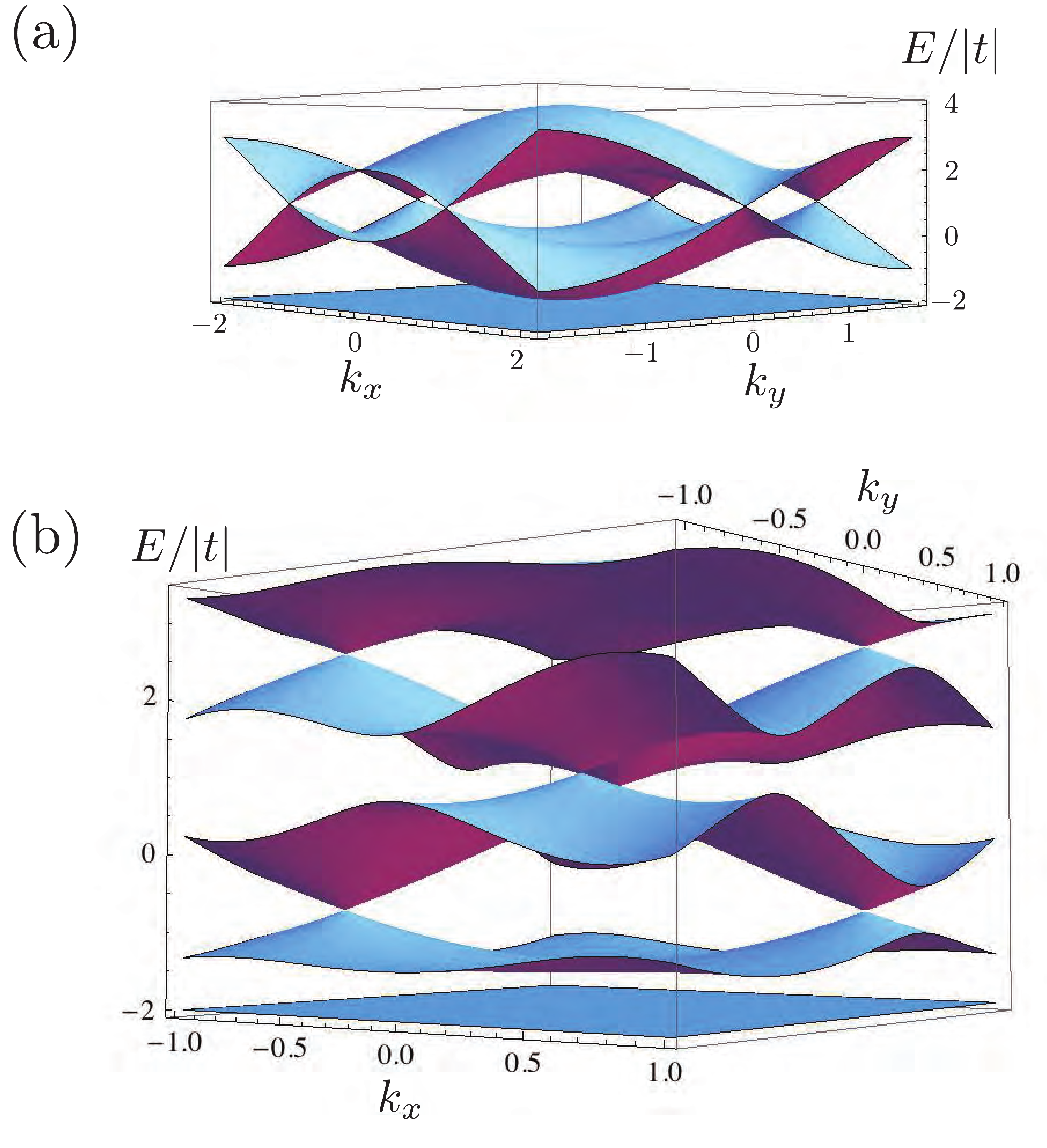}
%\centering
\caption{Band structure of kagome lattice tight-binding models at 0 and $\pi$ flux. (a) 0-flux model.  There is  a band touching of the lowest flat band with the higher dispersive band occurred at the $\Gamma$-point ($k_x=0, k_y=0$). (b) $\pi$-flux model. The total six bands in this model, corresponding to its six-site unit cell. Note that the lowest flat band(s) are doubly degenerate and gapped from the higher dispersive bands, with the minimum gap being $\Delta \approx 0.55 \abs{t}$.  }
\label{bandnew}
\end{figure}

We first discuss the tight-binding band structure for $U=0$.  In the absence of external flux, the single-particle spectrum has a lowest flat band as shown in Fig.~\ref{bandnew}(a).  However, the flat band touches the higher dispersive band at the $\Gamma$-point [$\mathbf{k}=(0,0)$]. 
Once flux $\phi$ is inserted into each hexagon of the kagome lattice, the single-particle spectrum $E(\phi)$ takes the typical Hofstadter butterfly form [Fig.\ \ref{butterfly}]. For non-integer $\phi/2\pi$, the lowest band \footnote{The term ``band" is applied loosely here. For rational flux, the unit cell is enlarged and the flat band decomposes into multiple ones. No such simple picture applies to irrational flux.} remains flat but acquires a gap that reaches its maximum size of $\Delta \approx  0.55\abs{t}$ at $\pi$ flux [Fig.~\ref{bandnew}(a)]. At this point, time-reversal (TR) symmetry is intact and the model can be realized with real-valued hopping of positive and negative sign.
The energy and degeneracy of the flat band are independent of the flux. The latter is given by the number of hexagons  in the lattice, $N_\text{site}/3$ ($N_\text{site}$ denoting the number of sites).  

\begin{figure*}[t]
\includegraphics[width=0.8\textwidth]{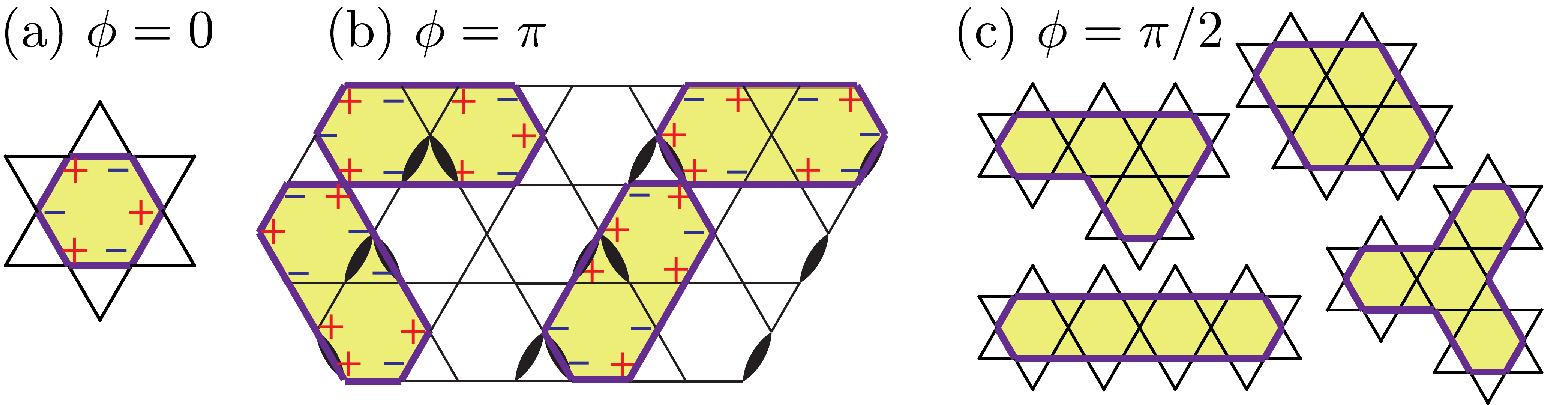}
\caption{(color online). (a)-(c) Maximally compact loop eigenstates for $\phi=0, \pi$ and $\pi/2$.  Thick bonds in (b) mark links with negative hopping $-|t|$. ``$\pm$" denotes the sign of the wavefunction amplitude on the corresponding sites.  Note that for (b) and (c), the sites in the interior of the loops have zero amplitude.}
\label{LoopSamples}
\end{figure*}

\subsection{Localized loop eigenstates}\label{lle}
The presence of the flat band is directly linked to the existence of degenerate eigenstates that form localized loops. The localization mechanism, called caging \cite{Vidal:1998ux, PhysRevB.91.041102}, is due to the destructive interference of the wavefunction amplitude anywhere outside the loop \cite{Bergman:2008ce}. In the kagome lattice with positive hopping, the flat band and localization persist as long as there is no flux through triangles. 

To be systematic, we list three important properties of the single particle eigenstates in the flat band of the tight-binding kagome Hamiltonian:
\begin{enumerate}
\item  The energy of the single-particle eigenstate is exactly -2$\abs{t}$.
\item    The elementary flat-band eigenstates are single-particle loop states that have equal probability on each involved site.  The amplitudes on the adjacent sites outside the loop (on the outward and inward triangles) cancel due to destructive interference  (caging). Any other (non-elementary) flat-band state can be composed as a linear superposition of loop states.
\item    ``Flux quantization":  any loop eigenstate encloses an integer number of flux quanta, 
\be
\phi_\text{L}=\otsum_\text{loop}\!A_{\mathbf{r}\mathbf{r}'} = \sum_\text{loop}\phi \in 2\pi\NN.
\ee
Here, the direction of the gauge potentials $A_{\mathbf{r}\mathbf{r}'}$ is chosen to be counterclockwise ($\circlearrowleft$) around the loop.
\end{enumerate}

Previous studies \cite{Bergman:2008ce, Zhitomirsky:2004ip,  Zhitomirsky:843638, Huber:2006bc} have focused on the zero-flux case and identified the state with amplitudes of equal magnitude but alternating signs on the hexagon loop  as maximally compact eigenstates  [see Fig.~\ref{LoopSamples}(a)].  While the existence of flat band and localized eigenstates remains unharmed by the flux through hexagons, we find that the shapes of loop states must change. Specifically, in the spirit of the flux-quantization condition (property 3), maximally compact loop eigenstates encircling two and four hexagons in Figs.~\ref{LoopSamples}(b) and \ref{LoopSamples}(c), respectively. Note that orientation and the shape of the maximally compact loop states are not generally unique for $\phi\not=0$.

The main focus of this paper is the case of flux $\phi=\pi$ where the TR symmetry is intact and maximally compact loop states are dimers encircling two hexagons. By a convenient gauge choice  $A_{rr'}=\pi$ on decorated bonds  [Fig.~\ref{butterfly}(b)], all hopping elements are real and given by $\abs{t}$ on regular and $-\abs{t}$ on decorated bonds. In this gauge, the amplitudes of loop eigenstates simply alternate in sign across positive-hopping bonds and are identical across decorated negative-hopping bonds.
In the following, we consider the occupation of these states by multiple bosons and refer to the maximally compact loop states as loop orbitals (LOs).

\section{Exact nematic Wigner crystal ground state}\label{exactsolution}
We next turn to the interacting case, accounting for on-site boson repulsion due to the Hubbard term 
\be
V=\sum_{\mathbf{r}} V_{\mathbf{r}}= \sum_{\mathbf{r}}  U b^\dag_{\mathbf{r}} b^\dag_{\mathbf{r}} b_{\mathbf{r}} b_{\mathbf{r}}.
\ee
Since the interaction is local, we note that \emph{any} many-body state of the form 
\be
\ket{\psi}=\prod_{m \in \mathcal{A}}   L^\dag_{m} \ket{0}
\ee
with single-particle occupation of a set $\mathcal{A}$ of \emph{non-overlapping} LOs is an exact ground state of the interacting system for filling $\nu=|\mathcal{A}|/N_\text{site}$ \footnote{Note: for filling below close packing, there will generally be a large number of degenerate ground states.}. Here, the operator $L^\dag_{m}$ creates a single particle occupying the LO labeled by $m$.  Indeed, the above product state is an eigenstate with eigenenergy $-2|t|$ per particle and interaction does not contribute since double occupancy of sites is avoided.   

\begin{figure}
\includegraphics[width=1\columnwidth]{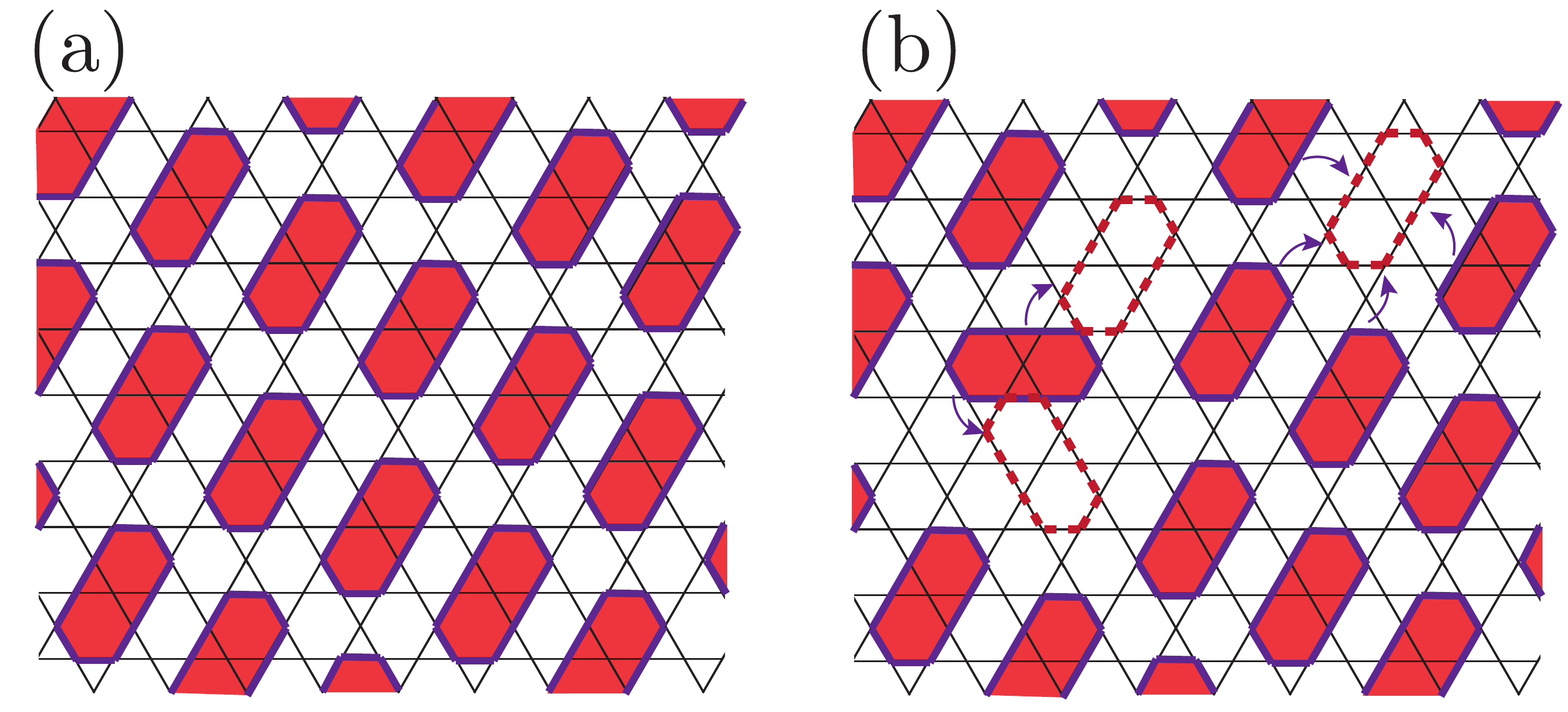}
\caption{(a) Illustration of a nematic Wigner crystal ground state, which breaks $C_6$ lattice rotational symmetry for $\phi=\pi$ and at close-packing filling $\nu_c=1/15$.    (b)  Illustration of the hard-core loop gas for $\phi=\pi$. The average density is slightly lower than the close-packing filling $\nu_c$.  }
\label{NematicWignerCrystal}
\end{figure}

Once the filling reaches close packing, the ground state becomes an incompressible Wigner crystal \cite{Wu:2007iz}. No additional particle can be placed on the lattice without incurring an interaction-induced energy increase due to unavoidable overlap. At the critical filling $\nu_c$ of close packing, bosons occupy maximally compact LOs while avoiding double occupation. As discussed above, maximally compact LOs may break the lattice point group symmetry (here, $D_6$), which directly leads to ground states with spontaneously broken lattice symmetry.

In general, the filling fraction $\nu_c$ for close packing depends on flux. In the $\pi$-flux case, maximally compact LOs are dimers and close packing occurs at $\nu_c\!=\!1/15$.  Due to the three possible orientations of a dimer [Fig.~\ref{butterfly}(b)] and the freedom to use one Wigner crystal representative [Fig.~\ref{butterfly}(d)] and produce four other inequivalent ones by translations to four neighboring hexagons, we predict that the  ground state is overall 15-fold degenerate. These ground states are nematic Wigner crystals. Here, nematicity refers to the emergence of dimers that break the lattice rotation symmetry. In this aspect, the $\pi$-flux case is dramatically different from the $0$-flux case studied before in the context of both the boson model and the antiferromagnetic Heisenberg spin model \footnote{Note that the Wigner crystal state, is first discovered in the spin context, and termed as valence-bond crystal. It is the exact ground state corresponding to the $m=7/9$ magnetization plateau \cite{Schulenburg:2002bv} of the antiferromagnetic kagome Heisenberg model in an external magnetic field. See Appendix \ref{spinmodel} for details.}, where the $\nu=1/9$ Wigner crystal  \cite{Huber:2006bc} and the $m=7/9$ valence-bond crystal \cite{Schulenburg:2002bv} ground states do not exhibit any nematicity.   We note that the nematic Wigner crystal state can also be found in the magnetization plateau of the corresponding spin model in the presence of both positive and negative XY interaction (see Appendix \ref{spinmodel}). 

We note that  for $\phi=\pi$ and filling below $\nu_c$, bosons form an infinitely compressible hard-core loop gas with macroscopic degeneracy determined by all possible configurations of non-overlapping loops \cite{Zhitomirsky:2004ip,  Zhitomirsky:843638}.  One such configuration is depicted in Fig.~\ref{NematicWignerCrystal}(b), where an extra particle can be added to an unoccupied loop orbital without costing any interaction energy. Hence the chemical potential, i.e.~energy cost per extra particle $\mu=\frac{\partial E}{\partial N}$, is fixed to be the flat-band energy $-2\abs{t}$ and hence does not change with the filling, i.e.~$\frac{\partial \mu}{\partial \nu} = 0$. Equivalently, we get $\frac{\partial \nu}{\partial \mu} = \infty$, which means infinite compressibility. As shown in Fig.~\ref{NematicWignerCrystal}(b), by adding a local perturbation, certain loops can move freely to a nearby vacancy (dashed loops) and hence make a transition to another state with the same ground-state energy.    The shown state (and other states which connect to this state by a local perturbation) breaks the lattice rotational symmetry and hence is also nematic. 
Finally, we mention that there are infinitely many incompressible glassy states below $\nu_c$, which cannot be connected to other ground states by a local perturbation. We will leave the discussion of these glassy states to future works.

\section{Nematic superfluid and supersolid}\label{nematic}

\subsection{Flat-band projection and construction of Wannier orbitals.}
In the following, we exclusively focus on the $\pi$-flux case. For filling above close packing, interaction cannot be avoided anymore and hence no exact solution in the above manner is possible. To make approximations, we derive a low-energy effective Hamiltonian by adapting the approach by Huber and Altman \cite{Huber:2006bc}, consisting of a projection onto the subspace spanned by flat-band eigenstates. In our case of nonzero flux, however, we forego the more subtle situation of an ungapped band encountered in Ref.~\cite{Huber:2006bc}.  In the presence of a gap and in the  weak-interaction limit,  boson occupation is to a good approximation limited to the flat band and the projection is appropriate unless the filling fraction becomes too large (details depend on the ratio $U/|t|$). 

\begin{figure*}
\includegraphics[width=1.5\columnwidth]{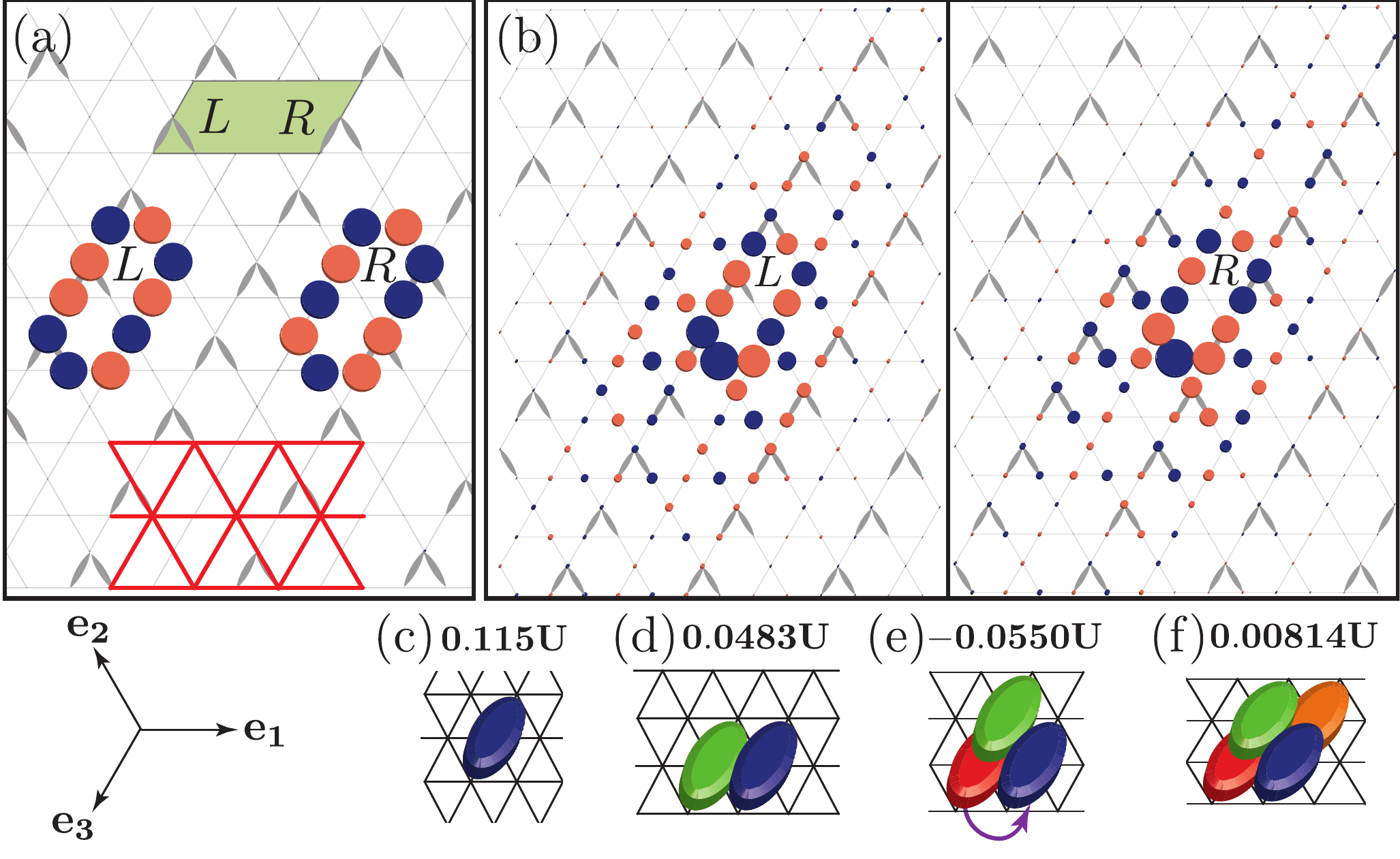}
\caption{(color online). (a)  The two types of dimer LOs used to construct the Wannier basis. Disks in orange/blue (light/dark gray) represent positive/negative amplitudes of the LO.  The shaded area on top shows a single unit cell,   containing two hexagons, $L$ and $R$.  The centers of LOs form a triangular lattice (bottom).  (b)~Two neighboring orthogonal Wannier orbitals. The area of each disk is proportional to the amplitude of the wavefunction on that site.   Bottom panels show four types of effective interactions: (c) on-site repulsion, (d) density-density repulsion,  (e) assisted-hopping, and (f) ring-exchange.}
\label{Wannier}
\end{figure*}

To facilitate the projection, we construct an orthonormal basis of the flat band. 
For $\pi$ flux, the unit cell is doubled and contains a left and right hexagon, $L$ and $R$, which differ by the relative positions of negative-hopping bonds [Fig.~\ref{Wannier}(a)]. Due to the unit-cell doubling there are, strictly speaking, two degenerate flat bands. Accordingly, we choose two sets of maximally compact dimer LOs aligned in the $\mathbf{e}_3$ direction [Fig.~\ref{Wannier}(a)] as our basis for the two degenerate flat bands. We distinguish left-dimer states only containing $L$ hexagons from right-dimer states only containing $R$ hexagons.
Although these sets of LOs together form a basis of the two degenerate flat bands, not all basis states are mutually orthogonal.  We thus need to determine appropriate superpositions of the dimer LOs to form a set of mutually-orthogonal Wannier orbitals (WOs). As usual, there is not a unique set of WOs and different choices can vary significantly in their real-space localization. Since we will ultimately employ local-decoupling mean-field theory, it is particularly important to obtain well-localized WOs \footnote{ Note that it is possible to construct $C_6$ symmetric but less compact localized orbitals as our flat-band basis. However, the mean-field ansatz with such orbitals has a much larger interaction energy cost due to the larger overlap and hence is not energetically favorable (see Appendix \ref{energetics} for details).}.

Our construction scheme for suitable WOs involves an important step of orthogonalizing the sets of left and right LOs by means of a symmetrized version of the Gram-Schmidt procedure (see Appendix \ref{construction} for details).  The results for two adjacent WOs are depicted in Fig.~\ref{Wannier}(b). The major part of the real-valued WO amplitude is essentially concentrated on each original dimer [Fig.~\ref{Wannier}(a)].  From there, the amplitudes decrease rapidly (asymptotically in an exponential fashion). This is in contrast to the slower power-law decay of WO amplitudes in the 0-flux case, which is caused by the touching of bands  \cite{Huber:2006bc}. The WOs we obtain respect translational symmetry (in terms of probability), TR symmetry, and preserve the mirror symmetry along their major axes, just as the original dimer LOs. They weakly break mirror symmetry along their minor axes.  

We define the creation operator for occupation of these Wannier orbitals by
\be\label{wannierdef}
\text{w}^{\dag}_{j} \equiv \sum_{\mathbf{r}}  w_j ({\mathbf{r}}) b^{\dag}_{\mathbf{r}} ,
\ee
where the Wannier function $w_j(\mathbf{r})$ gives the amplitudes of the dimer-type WO centered at position $j$ of the effective triangular lattice [Fig.~\ref{Wannier}(a)] on each site $r$ of the underlying kagome lattice. The flat-band projection corresponds to the inverse transformation
\be\label{proj}
b^\dag_{\mathbf{r}} \rightarrow \sum_j w_j^*(\mathbf{r}) \text{w}^\dag_j,
\ee
where the Wannier states of the dispersive bands have been dropped as an approximation.
Upon projection and switching to the grand-canonical ensemble, the effective Hamiltonian takes the form
\be\label{effectiveHamiltonian}
H \rightarrow H_\text{eff}= \sum_j (-2\abs{t}-\mu) \text{w}^\dagger_j   \text{w}_j + \sum_{i j k l} I_{ i j k l} \text{w}^\dagger_i \text{w}^\dagger_j \text{w}_k \text{w}_l,
\ee
where $\mu$ is the chemical potential.  For convenience, we may define the shifted chemical potential $\mu'=\mu+2\abs{t}$ which absorbs the energy constant of the flat band. The coefficients $I_{ijkl} \equiv U \sum_r w^*_i(r) w^*_j(r) w_k(r) w_l(r)$ determine the strength of the effective interaction terms and involve overlaps of four Wannier functions centered on specific sites $i$, $j$, $k$, and $l$ of the triangular lattice. Due to the localization of WOs, the interaction is short range and falls off rapidly with growing spatial distance between the four sites.  We note that $I_{ijkl}$ is translationally invariant and real-valued (since the constructed Wannier functions are real-valued themselves). 

The distinct spatial  configurations of the four dimer WOs labeled $i$ through $l$ give rise to different types of interaction terms. Whenever all four indices coincide,  the contribution corresponds to an effective onsite repulsion $  \sum_j U'\text{w}^\dagger_j \text{w}^\dagger_j \text{w}_j \text{w}_j$ with strength $U'=I_{jjjj}=0.11U$. Among the set of all effective interaction terms, this on-site repulsion term has the largest strength. The next sub-leading terms come from two other types of effective interaction, namely, density-density repulsion $V_\text{DD}$$=$$\sum'_{i, j} I^{\text{d}}_{ij} \text{n}_i \text{n}_j$ [Fig.~\ref{Wannier}(d)], and assisted hopping $V_\text{AH}=\sum'_{i,j} \sum_k I^{\text{ah}}_{ijk} \text{w}^\dag_i \text{w}_j \text{n}_k$  [Fig.~\ref{Wannier}(e)].  Here, $\text{n}_j $$\equiv $$\text{w}^\dag_j \text{w}_j $ denotes the Wannier number operator, and the primes on sums signal that those terms with coinciding summation indices are to be omitted.  The strengths of density-density interaction and assisted hopping depend on the specific arrangement of the involved WO dimers. The largest contributing terms are $I^\text{d}_{\mathbf{0},\mathbf{e_1}}$$=$$0.0483U$ and $I^\text{ah}_{\mathbf{0},\mathbf{e_1}, -\mathbf{e_3}}$$=$$-0.055U$,  and thus significantly smaller than the on-site repulsion strength $U'$. 

Within the low-density regime $\nu$$<$$1/3$ (i.e., $\nu_\text{eff} $$<$$1$ in the effective triangular lattice), we therefore employ a hard-core approximation which forbids double occupation of WOs \cite{Huber:2006bc}.  
%Thus,  the Wannier operator $\text{w}_j$ may be replaced by the Pauli operator $\sigma_j^-$.
Within this approximation, interaction terms with repeated Wannier operators on the same site, $\text{w}^\dagger_j\text{w}^\dagger_j$ or $\text{w}_j\text{w}_j$, drop out. This includes effective onsite interaction as well as pair hopping $\sum'_{i,j} I_{ iijj} \text{w}^\dagger_i \text{w}^\dagger_i \text{w}_j \text{w}_j$. Besides density-density repulsion and assisted hopping, the only remaining interaction type is ring-exchange [Fig.~\ref{Wannier}(f)], in which the Wannier functions are centered on four different sites on the triangular lattice. We find that the maximum strength of ring exchange is $0.00814U$, which is significantly weaker than both density-density repulsion and assisted hopping.

\subsection{Mean-field theory.}\label{MFT}

While density-density repulsion favors density-wave order and formation of a Wigner crystal, assisted hopping may lead to melting and formation of a superfluid. In addition, this competition also allows for an intermediate supersolid phase in which both types of order are present. Here, we study the competition between different types of orders within mean-field theory (MFT). We adopt the Gutzwiller approach \cite{Rokhsar:1991vo} and employ a product ansatz consistent with the hard-core constraint  
\be\label{wavefunction}\textstyle
\ket{\psi_\text{MF}}=\prod_j  (f_{j, 0} +  f_{j, 1}   \text{w}^\dagger_j) \ket{0}, 
\ee
which decouples sites on the effective triangular lattice of WOs. 
The mean-field ansatz naturally captures the nematic Wigner crystal phase since it is a product of single-particle states with occupation of non-overlapping LOs (in this case approximated by WOs). Above close packing, mean-field solutions continue to break the $C_6$ symmetry due to the anisotropic nature of the Wannier orbitals. 
%We employ the hard-core approximation discussed above, thus limiting WO occupation to $n$$=$$0, 1$. We note that coherent superposition of different number states (e.g.~ $n=0,1$) suggests the presence of superfluidity. 

To describe states with density-wave order such as the nematic Wigner crystal, we must allow for the dependence of the mean-field amplitudes $ f_{j,n}$ on the spatial index $j$. To obtain 
 mean-field solutions, we decouple the effective Hamiltonian, replacing density-density interaction and assisted-hopping terms by 
 \begin{align}\label{decoupling}
\nonumber V_\text{DD} \rightarrow& {\sum_{i, j}}'  2 I^{\text{d}}_{ij} \text{n}_i  \langle \text{n}_j \rangle,   \\
\nonumber V_\text{AH}   \rightarrow &  {\sum_{i,j}}'\sum_k  I^{\text{ah}}_{ijk}   \langle  \text{w}^\dagger_i \rangle \langle  \text{w}_j \rangle   \text{n}_k \\
+& {\sum_{i,j}}' \sum_k I^{\text{ah}}_{ijk} (  \text{w}^\dagger_i  \langle  \text{w}_{j} \rangle +     \text{h.c.}) \langle \text{n}_{k} \rangle.
 \end{align}
(We have verified that inclusion of ring-exchange does not lead to significant changes.) With this, we obtain a mean-field Hamiltonian $\sum_i h_i (\{\psi_j \}, \{n_j\} )$, where $h_i$ depends on the mean-field order parameters  $\psi_j= \langle \text{w}_j \rangle $ and $n_j=\langle \text{n}_j \rangle$ on each site of the triangular lattice. Starting from a random initial set of order parameters on a  lattice of 200 sites with periodic boundary conditions, we repeatedly solve for the eigenstates and re-calculate order parameters until reaching self-consistency (see Appendix \ref{algorithm} for details). 

\begin{figure}
\includegraphics[width=1\columnwidth]{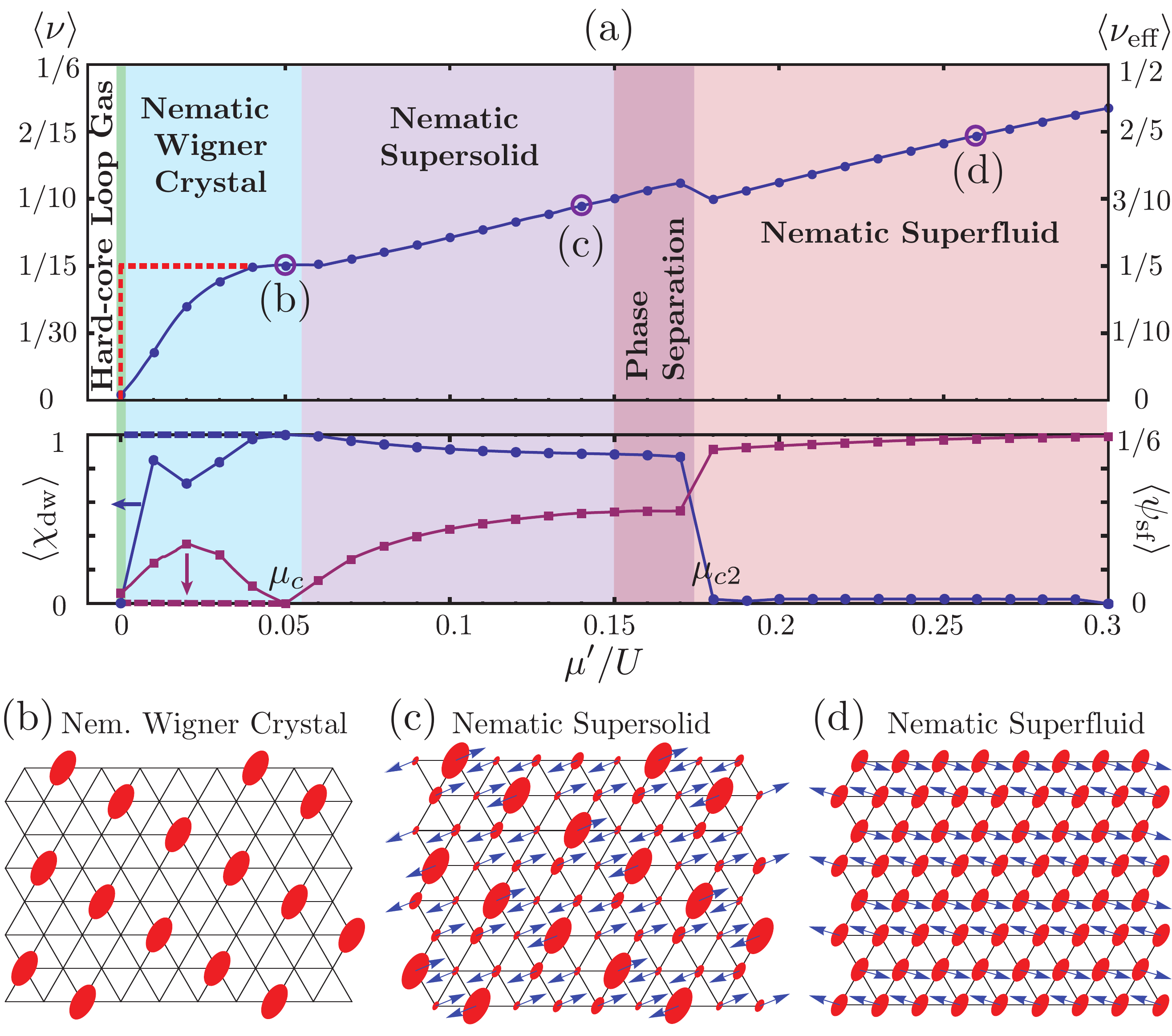}
\caption{(color online). (a) Mean-field phase diagram and plot of average density $\langle \nu \rangle$, $\langle \nu_\text{eff} \rangle$ on the effective triangular lattice, average density-wave order $\langle \chi_\text{dw} \rangle$ (blue dots) and average superfluid order $\langle \psi_\text{sf}  \rangle$ (red squares) versus chemical potential.  Dashed lines show the exact solution with LOs.  (b-d) Results from MFT for ground states in the three phases. The area of depicted dimers is proportional to the local occupation number $n_i$, arrows show the phase angle $\theta_i$. Results are obtained for a lattice of 200 sites using a  self-consistency calculation.}
\label{phasediagram}
\end{figure}

For a range of chemical potentials, we calculate  results for the mean filling $\langle \nu \rangle   \equiv \sum_i n_i /N_\text{site}$, density-wave order parameter $\langle \chi_\text{dw} \rangle$ defined as the difference between maximum and average density taking into account the six surrounding sites,  and the mean superfluid order parameter $\langle \psi_\text{sf} \rangle$$\equiv$$\sum_i \psi_i /N_\text{site}$. The key results from this calculation are presented in Fig.~\ref{phasediagram}. MFT reproduces the exact nematic Wigner crystal [Fig.~\ref{phasediagram}(b)] for $\mu'\simeq0.05 U$  at close packing $\langle \nu \rangle=1/15$, showing maximum density-wave order $\langle \chi_\text{dw} \rangle=1$ and vanishing superfluidity $\langle \psi_\text{sf} \rangle=0$.   Below close packing, MFT produces a gradual change of average filling and superfluid order, which differs from the exact solution discussed above based on LOs.  The exact solution exhibits a density plateau at $\nu_c$ containing the entire nematic Wigner crystal phase, and a vertical jump corresponding to the hard-core loop gas phase which is more appropriately represented in the canonical ensemble.

Above $\mu_c$, superfluid order sets in and $\langle \psi_\text{sf} \rangle$ grows gradually while, at the same time, the density-wave order parameter $\langle \chi_\text{dw} \rangle$ remains nonzero and decays slowly,  overall suggesting a second-order transition to a nematic supersolid [Fig.~\ref{phasediagram}(c)] in which a fraction of the bosons condense on  interstitial sites between the Wigner-crystal structure. Further on at $\mu_{c2}$$\simeq$$0.18 U$, the density-wave order $\langle \chi_\text{dw} \rangle$ abruptly drops to zero, accompanied by a sudden increase in the superfluid order $\langle \psi_\text{sf} \rangle$. This indicates a sudden melting of the Wigner-crystal structure and a first-order transition into a  superfluid phase [Fig.~\ref{phasediagram}(d)].     

Based on our MFT, we predict that the superfluid phase is nematic since condensation of bosons is based on hopping among anisotropic dimer WOs. Within the superfluid phase, phase angles $\theta_i=\text{Arg}[\psi_i]$ form stripes in which neighboring stripes differ by a $\pi$-phase difference.      The nematic supersolid has similar phase stripes, the only difference being that sites with maximum density have an additional $\pi$-phase flip.   
Finally, we find a narrow region in which the non-monotonic dependence of the density on the chemical potential suggests phase coexistence between the  superfluid and supersolid.

\section{The signatures of nematic superfluidity and detection methods}\label{signature}

The interesting aspect of the uniform superfluid phase is that it is nematic. Its internal structure, i.e., the correlation in the loop/Wannier orbitals, is encoded in the
momentum distribution, which can be probed through time-of-flight (TOF) imaging in the context of cold-atom experiments. The nematicity can also be identified through macroscopic
quantities, such as the superfluid stiffness tensor and the anisotropic superflow, which can be probed with phase imprinting technique \cite{Isoshima:2000vh}.  In the following two sections, we discuss microscopic and macroscopic signatures, along with methods to detect them.

\begin{figure}
\includegraphics[width=1\columnwidth]{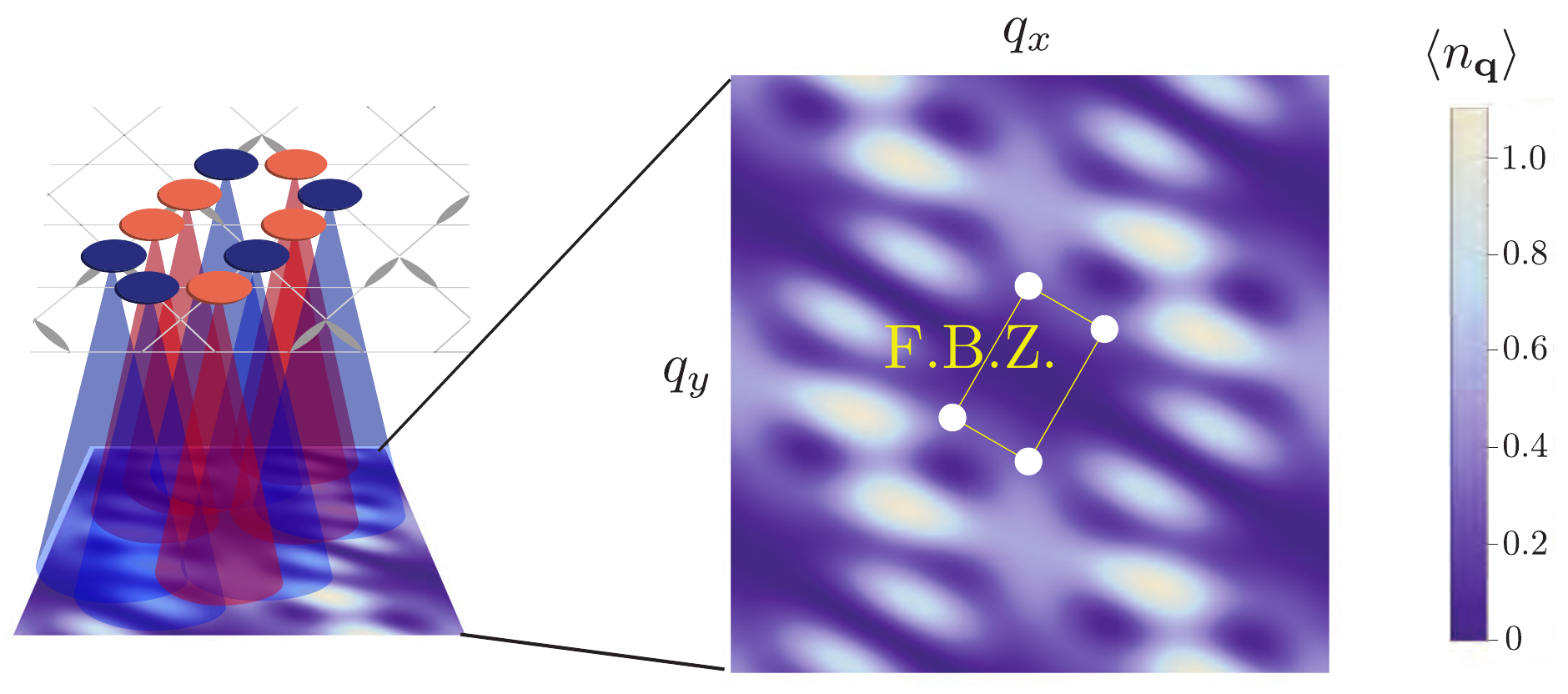}
\caption{(color online). TOF image probing the momentum distribution of the superfluid phase.   Besides the anisotropic background revealing the internal correlations of loop orbitals, the delta function peaks correspond to the condensate in $k$-space. Only the four peaks (corresponding to equivalent points in the reciprocal space) on the first-Brillouin zone boundary (yellow box) are shown.}
\label{TOF}
\end{figure}

\begin{figure*}
\includegraphics[width=2 \columnwidth]{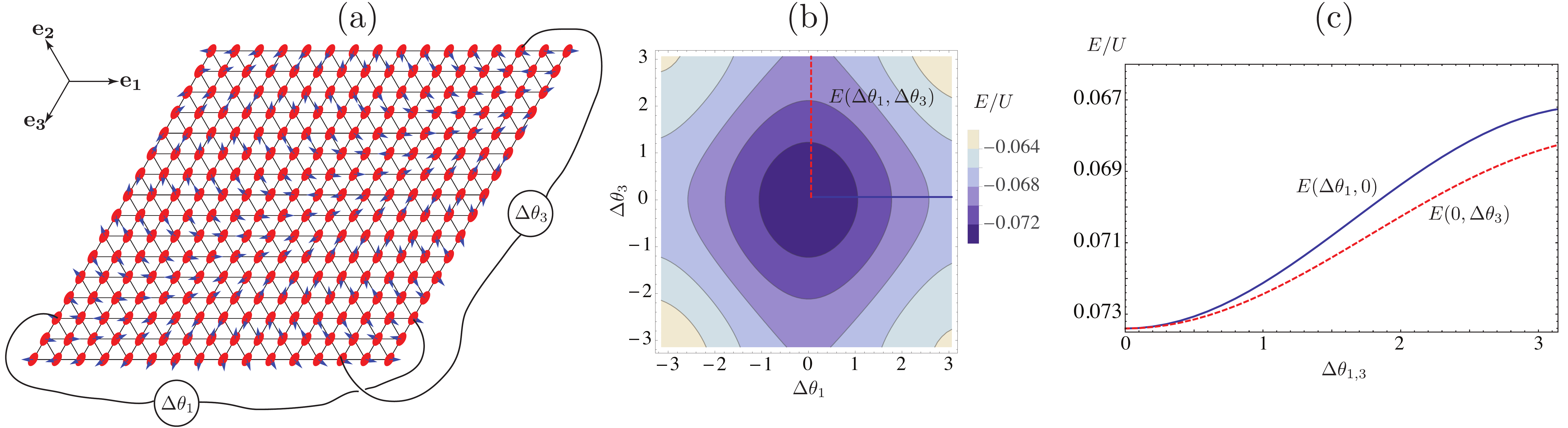}
\caption{(color online). (a) Applying phase differences $\Delta \theta_1$ (in the $\mathbf{e}_1$ direction) and $\Delta \theta_3$ (in the $\mathbf{e}_3$ direction) across the boundaries of the finite sample (16 $\times$ 16).  The configuration of phase angles illustrates the situation of $\Delta \theta_1 = \Delta \theta_3 = \pi$.  (b) Contour plot showing the mean-field energy distribution as a function of the phase differences, i.e., $E(\Delta \theta_1, \Delta\theta_3)$.  The energy profiles along the two cuts (blue solid and red dashed lines),  $E(\Delta \theta_1, 0)$ and $E(0, \Delta \theta_3)$, are shown in (c). }
\label{stiffness}
\end{figure*}

\subsection{Momentum distribution and time-of-flight experiments}\label{momTOF}

The microscopic signature, i.e., the ground-state momentum distribution $\langle n_\mathbf{q} \rangle$, as mentioned above, can be directly measured experimentally through the time of flight images. It serves as a useful probe of the correlation properties of the ground state.    The momentum distribution is equal to the Fourier transform of the single-particle density matrix:
\be
\langle n_\mathbf{q} \rangle= \frac{1}{N_\text{site}} \sum_{\mathbf{r}_1, \mathbf{r}_2} e^{i \mathbf{q} \cdot (\mathbf{r}_1-\mathbf{r}_2)} \langle b^\dag_{\mathbf{r}_1}  b_{\mathbf{r}_2}  \rangle.
\ee

Recall that for the nematic Wigner crystal (NWC) ground state, the wavefunction can be expressed as $\ket{\text{NWC}}$$=$$\prod_{j \in \mathcal{A}} L^\dag_j \ket{0}$.   Here, $j$ is the label of the loops and $\mathcal{A}$ is the set of non-overlapping loop states forming the crystal.  Since the loops do not overlap, we have the correlation $ \langle b^\dag_{\mathbf{r}_1}  b_{\mathbf{r}_2}  \rangle \neq 0$ if and only if $\mathbf{r}_1$ and $\mathbf{r}_2$ are sites on the same loop.  Thus, we get
\be\label{NWCcorrelation}
\langle n_\mathbf{q} \rangle_\text{NWC} = \frac{1}{N_\text{site}}  \sum_{j \in \mathcal{A}} \sum_{\alpha, \alpha'} e^{i \mathbf{q}\cdot (\mathbf{r}_\alpha-\mathbf{r}_{\alpha'})} \bra{0} L_j b^\dag_\alpha  b_{\alpha'} L^\dag_j \ket{0},
\ee
where $b^\dag_\alpha$ represents the boson operator on the loop ($\alpha= 1,2,3,...,10$).    There are two types of loops: one encircles the left hexagons and the other encircles the right hexagons, as shown in Fig.~\ref{Wannier}(a).  As we can see, the nonzero contributions for the single-particle density matrix in Eq.~(\ref{NWCcorrelation}) are $ \pm \bra{0} \frac{1}{10} b_\alpha b^\dag_\alpha b_{\alpha'} b^\dag_{\alpha'}  \ket{0}= \pm \frac{1}{10}$, where ``$\pm$" is determined by the sign of the overlap of wavefunction amplitudes.   
 
Now we discuss the momentum distribution of the nematic superfluid phase (see Appendix \ref{TOFcalculation} for derivation).  It is shown in Fig.~\ref{TOF}. The continuous background originates from the correlation within the loop orbitals; apart from a prefactor it is identical to NWC.  It is squeezed in the direction
of the major axis (along which the loop is elongated).  This continuous background encodes the internal correlation of the loop/Wannier orbitals.  The delta-function peaks (represented by white circles) originate from the Bose condensation. Only the four equivalent peaks on the boundary of the first Brillouin zone are shown in the plot. Such a momentum distribution of a uniform superfluid implies a novel scenario of Bose condensation, where the ground state is unstable against developing an additional nematic order which, spontaneously breaks the lattice rotational symmetry. Therefore in the corresponding TOF experiment, one expects that the sample prepared under similar conditions repeatedly would spontaneously pick up one of the three directions in which the image pattern is squeezed.

\subsection{Superfluid stiffness tensor and anisotropic superflow}

Now we discuss the macroscopic signature of the nematic superfluid phase. We consider the superfluid stiffness tensor $\rho_{IJ}$$=$$ \frac{\partial^2 E}{\partial \Delta \theta_I  \partial  \Delta \theta_J }|_{\Delta \theta_{I,J}=0}$, where $I, J=1,3$ refer to the directions along $\mathbf{e}_1$ and $\mathbf{e}_3$.  Note that $\mathbf{e}_3$ direction is special because it aligns with the major axis of the dimers.
To study this quantity, we apply phase differences $\Delta \theta_1$ and $\Delta \theta_3$ across the boundaries of the finite sample (16 $\times$ 16), as shown in Fig.~\ref{stiffness}(a).     Experimentally such phase difference can be achieved with the phase imprinting technique developed in the cold atom setup \cite{Isoshima:2000vh}.

The phase differences across the boundaries induce superflow in the corresponding directions and hence increase the kinetic energy.    The contour plot in panel (b) shows the mean-field energy as a function of the phase differences, namely $E(\Delta \theta_1, \Delta\theta_3)$.  The superfluid stiffness tensor corresponds to the curvature of the energy profile in the vicinity of the origin.   The anisotropy of the energy contours suggests that the superfluid stiffness is also anisotropic.   To see this more clearly, we make cuts along the $x$- and $y$-axis (blue solid and red dashed line). We then show the energy profile along the two cuts, namely $E(\Delta \theta_1, 0)$ and $E(0, \Delta \theta_3)$, in panel (c).    It is obvious that, in the vicinity of the origin,  the curvature of the blue solid line is larger than that of the red dashed line, which means that $\rho_{11} \equiv \frac{\partial^2 E}{\partial \Delta \theta_1^2  }|_{\Delta \theta_1=0}$ is larger than $\rho_{33} \equiv \frac{\partial^2 E}{\partial \Delta \theta_3^2  }|_{\Delta \theta_3=0}$.  This suggests that the superfluid stiffness along the two directions is different.   Now we consider the first derivative,  $j_I \equiv \frac{\partial E}{ \partial \Delta \theta_I}$, which is the current generated when applying a phase difference at direction $I$.  We can see that, not far away from the origin, $j_1$ is always larger than $j_3$. This is consistent with the results in Appendix \ref{summaryeffective}, where we show that the effective nearest-neighbor hopping along the major axis ($\mathbf{e}_3$ direction) is zero in the nematic superfluid phase (within mean-field approximation).   Only successive hopping along other directions  will contribute to superflow in the $\mathbf{e}_3$ direction.   On the other hand, the large effective nearest-neighbor hopping in the other two directions ($\mathbf{e}_1$ and $\mathbf{e}_2$) leads to larger superflow in those directions.

In sum, the anisotropy of the two macroscopic quantities, the superfluid stiffness and superflow, reveals breaking of discrete rotational symmetry and hence  microscopic nematicity of the superfluid phase.

\section{Conclusion}\label{conc}
We studied the emergence of nematic phases in a kagome lattice with a gapped flat band, obtained when a flux $\pi$ is threaded through each hexagon of the lattice. Single-particle localized loop states can be combined to construct many-body eigenstates below a critical filling. This critical filling corresponds to close packing of nonoverlapping loop states and marks the formation of a nematic Wigner crystal ground state. For larger filling, the effective Hamiltonian based on flat-band projection using dimer-shaped Wannier orbitals and subsequent mean-field treatment predict nematic supersolid and superfluid phases.  The latter is a uniform quantum liquid with anisotropic internal structure, which is encoded in the momentum distribution and can be probed by time-of-flight experiment. Interesting future directions include the study of phases at higher density, especially the possibility of a featureless Mott insulator \cite{Parameswaran:2013dx} at 1/3 filling, resonating valence bond states and fermionized ground states in the strong-interaction regime.

%%%%%%%%%%%%%%%%%%%%%%%%%%%%%%%%%

\begin{acknowledgments}
We are indebted to Steven Girvin, Ashvin Vishwanath, Tigran Sedrakayan,  Eliot Kapit, Hakan T\"ureci, Tzu-Chieh Wei, Anupam Garg, James Sauls, Murad Tovmasyan and Andy C.~Y.~Li for insightful discussions.  Work performed at Northwestern University (G.Z. and J.K.) was supported by the NSF under Grant PHY-1055993. Work performed at Argonne National Laboratory (I.M.) is supported by the U. S. Department of Energy, Office of Science, Office of Basic Energy Sciences, under Contract No. DE-AC02-06CH11357.
\end{acknowledgments}

\begin{appendix}

\section{Derivation of the properties of loop eigenstates}\label{newproperty}
In this appendix, we show the derivation of some properties of loop eigenstates stated in Sec.~\ref{lle}. 

\begin{figure}
\includegraphics[width=0.7\columnwidth]{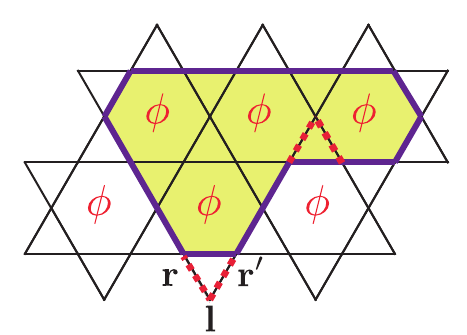}
\caption[Illustration of a general single-particle loop eigenstate]{Illustration of a generic single-particle loop eigenstate, where $\mathbf{r}$ and $\mathbf{r}'$ label sites on the loop, and $\mathbf{l}$ labels a site outside the loop where destructive interference (illustrated by red dashed lines) occurs.  }
\label{generalloop}
\end{figure}

The flux quantization condition (property 3) can be derived from properties 1 and 2 as follows.   The positive-hopping tight-binding lattice Hamiltonian in the presence of an additional gauge field $A$ has the form:
\be\label{tight-binding}
H_\text{tb}= \sum_{\langle \mathbf{r},\mathbf{r}'  \rangle} \abs{t} e^{i A_{\mathbf{r}\mathbf{r}'}} b^\dag_{\mathbf{r}'} b_{\mathbf{r}} + \text{H.c.} \equiv \sum_{\langle \mathbf{r},\mathbf{r}'  \rangle}T_{\mathbf{r}\mathbf{r}'},
\ee
where $T_{\mathbf{r}\mathbf{r}'}$ is the hopping operator on the nearest-neighbor bond $\langle \mathbf{r},\mathbf{r}' \rangle$.    Assuming that there exist single-particle wavefunctions of the loop eigenstate type, they can be expressed as  
 \be
 \ket{\psi_L} = \sum_{\mathbf{r}  \in L} \psi_{\mathbf{r}} b^\dag_{\mathbf{r}} \ket{0},
 \ee
 where the summation index $\mathbf{r}$ runs over all sites on the loop $L$.   When acting with the Hamiltonian $H_\text{tb}$ on the loop eigenstates,  we can split the expressions into two parts, namely
\be\label{split}
H_\text{tb} \ket{\psi_L} =\sum_{ \langle \mathbf{r},\mathbf{r}'  \rangle \in L} T_{\mathbf{r}\mathbf{r}'} \ket{\psi_L} + \sum_{ \langle \mathbf{r},\mathbf{r}' \rangle \in L,  \mathbf{l} \notin L } [ T_{\mathbf{r}\mathbf{l}} + T_{\mathbf{r}'\mathbf{l}}  ] \ket{\psi_L}.
\ee
The first sum includes hopping along the bonds $\langle \mathbf{r},\mathbf{r}' \rangle$ on the loop $L$,  while the second sum corresponds to hopping from the bonds $\langle \mathbf{r},\mathbf{r}' \rangle$ on the loop to the adjacent sites $\mathbf{l}$ on the outward/inward triangles, as illustrated in Fig.~\ref{generalloop} by red dashed lines.      The cancellation of the probability amplitude outside the loop (caging, property~2) requires the second sum to be zero (see Appendix \ref{newproperty}),  while the requirement of eigenenergy being $-2\abs{t}$ (property 1) implies the first sum being equal to:
\begin{align}
& \sum_{\langle \mathbf{r},\mathbf{r}' \rangle \in L} \abs{t} [e^{i A_{\mathbf{r}\mathbf{r}'}} \psi_\mathbf{r} b^\dag_{\mathbf{r}'} + e^{- i A_{\mathbf{r}\mathbf{r}'}} \psi_{\mathbf{r}'}b^\dag_\mathbf{r}] \ket{0}  \\
\nonumber &=-2\abs{t} \ket{\psi_L}   \equiv  \sum_{\langle \mathbf{r},\mathbf{r}' \rangle\in L} -\abs{t} [\psi_{\mathbf{r}'} b^\dag_{\mathbf{r}'} + \psi_\mathbf{r} b^\dag_\mathbf{r}] \ket{0}.
\end{align}
The above equation leads to the following relation between the amplitudes of neighboring sites,
\be\label{chain}
\psi_{\mathbf{r}'}=-\psi_\mathbf{r} e^{i A_{\mathbf{r}\mathbf{r}'}}.
\ee
That is, the wavefunction has equal probability on every site along the loop, and adjacent sites differ by a minus sign and an additional phase shift due to the gauge potential $A_{\mathbf{r}\mathbf{r}'}$ on the bond $\langle \mathbf{r},\mathbf{r}' \rangle$.  Note that in the 0-flux case ($A_{\mathbf{r}\mathbf{r}'}=0$),  Eq.~\eqref{chain} simplifies to alternating signs on the loop, including the hexagon loop state shown in Fig.~\ref{LoopSamples}(a).  By applying Eq.~(\ref{chain})  around the loop and requiring the probability amplitude $\psi_\mathbf{r}$ to be single-valued, we derive the flux quantization condition for a loop eigenstate (property 3), namely 
\be\label{flux1}
\phi_\text{L}=\displaystyle {\sum_{\langle \mathbf{r},\mathbf{r}' \rangle}\!\!\!\!\!\!\!\!\!\!\circlearrowleft } \ \ \  A_{\mathbf{r}\mathbf{r}'}=~ 2\pi n,  \   n\in \mathbb{N}.
\ee

Now we consider the cancellation of the second sum in Eq.~(\ref{split}) which leads to caging, and get 
\begin{align}
 \nonumber & [T_{\mathbf{r}\mathbf{l}}+T_{\mathbf{r}'\mathbf{l}}] \ket{\psi_L}  \\
 =& \abs{t}[e^{i A_{\mathbf{r}\mathbf{l}}} \psi_\mathbf{r} b^\dag_\mathbf{l} + e^{-i A_{\mathbf{r}'\mathbf{l}}} \psi_{\mathbf{r}'} b^\dag_\mathbf{l} +\text{H.c.} ] \ket{0} =0.
\end{align}
This in turn leads to $\psi_{\mathbf{r}'}=-\psi_\mathbf{r} e^{i (A_{\mathbf{r}\mathbf{l}}+A_{\mathbf{l}\mathbf{r}'})}$.   Combined with Eq.~(\ref{chain}), we get 
\be\label{flux2}
\phi_{\Delta}=A_{\mathbf{r}\mathbf{l}} + A_{\mathbf{l}\mathbf{r}'}+A_{\mathbf{r}'\mathbf{r}}=0.
\ee
Thus, we have just shown the necessary condition for the persistence of lowest flat band with energy $-2\abs{t}$, which we stated in the beginning of Sec.~\ref{lle}, is that the flux threading all the outward/inward triangles must be zero (gauge-equivalent to $\pi$-flux in the negative hopping model).

To understand one additional feature of the loop state, now we consider the gauge-invariant current operator on the bond $\langle \mathbf{r},\mathbf{r}'\rangle$ (from site $\mathbf{r}$ to site~$\mathbf{r}'$), namely,
\be
J_{\mathbf{r}\mathbf{r}'}=2 \abs{t}  i [ \ b^\dag_{\mathbf{r}'} b_{\mathbf{r}} e^{i A_{\mathbf{r}\mathbf{r}'}} - \text{H.c.}]. 
\ee
Its expectation value is given by
\be
\bra{\psi_L} J_{\mathbf{r}\mathbf{r}'} \ket{\psi_L}=2 \abs{t}  i [\psi^*_{\mathbf{r}'}  \psi_{\mathbf{r}} e^{i A_{\mathbf{r}\mathbf{r}'}} - \text{H.c.}],
\ee
 which equals zero after applying Eq.~(\ref{chain}).    Therefore, for any flux value $\phi$,  the loop states in the flat band carry no current, and thus do not break time-reversal (TR) symmetry, even though the Hamiltonian itself breaks TR except at $\phi=0, \pi$.   From the butterfly spectrum in Fig.~\ref{butterfly}(b), we observe that the lowest flat band does not change as a function of $\phi$, which leads to zero current due to the linear response formula of the current $J=\frac{\partial E}{ \partial \phi} =0 $.   The fact that there is no current can also be understood in another way, i.e.~the Chern number for the lowest flat band is always zero.  This is in contrast to the higher bands near $\phi=0$ [Fig.~\ref{butterfly}(b)], which are essentially Landau levels with nonzero Chern numbers $C=1,2,3,$ etc..

\section{Connection to frustrated spin models}\label{spinmodel}
Although the current paper focuses on the interacting boson models, some of the results,  such as the Wigner crystal phases, also apply for the frustrated spin models corresponding to the same type of frustrated lattice.  Here, we consider the frustrated anisotropic Heisenberg model with spin-$s$ in the presence of external magnetic field: 
\be
H= \sum_{\langle ij \rangle} [ J^{\perp}_{ij}  ( S^x_i   S^x_j + S^y_i  S^y_j  ) +  J^{z}_{ij}  S^z_i S^z_j    ] - h \sum_j S^z_j.
\ee
Here, $J^{\perp}_{ij}$ determines the $XY$ interaction and $ J^{z}_{ij}$ determines the $zz$ (Ising) interaction. The isotropic situation ($J^{\perp}_{ij}=J^{z}_{ij}$)  corresponds to the usual Heisenberg model.  The external magnetic field $h$ acts as the chemical potential of magnons and has nothing to do with the gauge flux which we discuss throughout the whole paper.   To make the connection with the boson model more explicit, we rewrite it as: 
\be
H= \sum_{\langle ij \rangle} [  \frac{1}{2} J^{\perp}_{ij}  ( S^+_i   S^-_j + S^-_i  S^+_j  ) + J^{z}_{ij}  S^z_i S^z_j    ] - h \sum_j S^z_j.
\ee
From above, one can see that the transverse $XY$ interaction can be written as a flip-flop term, which in the $s=\frac{1}{2}$ case actually corresponds to hopping of hard-core bosons. The $zz$ term induces nearest neighbor interactions between hard-core bosons.  If the magnetic field $h$ is sufficiently large, the ground state is the magnon vacuum $\ket{0} $$=$$ \ket{\uparrow\uparrow\uparrow\cdots\uparrow\uparrow\uparrow}$.    For the flat-band hopping models, a single magnon on the loop $m$ is created by the operator $L^\dag_m = \sum_{j \in m} \psi_j S^-_j$ out of the magnon vacuum $\ket{0}$,  where $\psi_j$ represents the wavefunction amplitude on each site of the loop.  Just like in the case of on-site interacting bosons, the magnon loop gas and the loop crystal are the eigenstates of the Hamiltonian (the $zz$ term does not change that since the adjacent loops do not occupy neighboring sites.  

Indeed, for the antiferromagnetic Heisenberg model ($J^{\perp}_{ij}=J^{z}_{ij}>0$), a valence-bond crystal phase (equivalent to the Wigner crystal) formed by non-overlapping hexagon loop magons has been found in Ref.~\cite{Schulenburg:2002bv} (earlier than its boson analog).   This valance-bond crystal phase, in the spin-1/2 case, corresponds to the $m=7/9$ magnetization plateau (m=0 corresponds to no polarization; $m=\pm1$ corresponds to full polarization in the up/down direction), and is equivalent to the 1/9 state of the interacting boson model.   We note that the valence-bond crystal phase is not limited to spin-1/2 case, but actually exists for arbitrary spin $s$ \cite{Schulenburg:2002bv}.

To implement the $\pi$-flux model described in the main text, one can choose different signs of the transverse coupling $J_{ij}$ according to the hopping signs of the corresponding boson model.   Thus, there will be a nematic valence-bond crystal (at 13/15 magnetization plateau) in the spin model which corresponds to  1/15 nematic Wigner crystal in the boson model.   To make sure the density-density interaction between magnons is repulsive and hence stabilizes the valence-bond crystal as the ground state, it is preferable that the $zz$-interaction is positive, namely $J^z_{ij} > 0$.  However, we caution that, if the anisotropy ratio $\abs{J^z_{ij}/J^\perp_{ij}}$ is sufficiently small, the sign of the $zz$-term does not matter since the $zz$-term can be treated as a small perturbation.  To experimentally realize such a sign-tunable spin model, a promising candidate is the nitrogen-vacancy center array \cite{Cai:2013gd}, although in that case the spin-spin interaction is not restricted to nearest-neighbors but has a power-law decay due to its dipole-dipole nature.

Besides the nematic Wigner crystal (nematic valence-bond crystal) phase, the nematic supersolid or superfluid phases may also occur in the spin systems.     However, due to the hard-core nature of the spin system (especially for spin 1/2), the projection approach may break down at certain critical filling.   Only below that critical filling, the predicted phase for the weakly-interacting boson model is expected to apply.

\section{Construction of orthogonal Wannier Orbitals}\label{construction}
\begin{figure*}
\includegraphics[width=2\columnwidth]{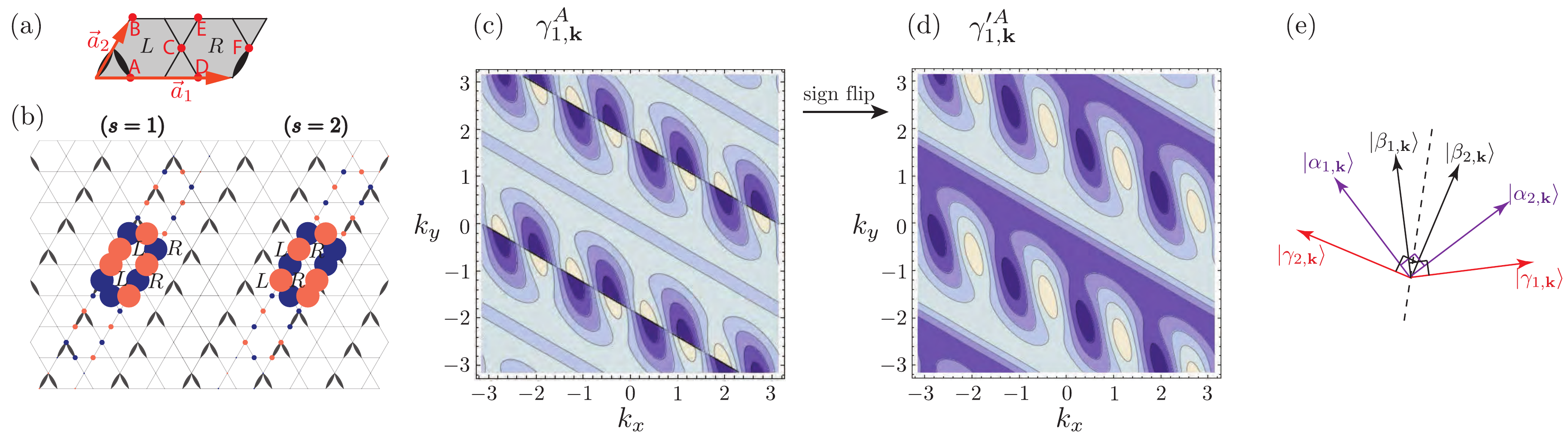}
\caption{(color online). (a) The unit cell of the $\pi$-flux kagome lattice, which contains two hexagons and six sites.  (b) Illustration of two Wannier functions transformed from the two Bloch vectors $\vec{\beta}_{1, \mathbf{k}}$ and $\vec{\beta}_{2, \mathbf{k}}$ via Eq.~(\ref{transformfunction}).   
   (c) The contour plot which shows the distribution of the A-component of the Bloch vector $\vec{\gamma}_{1,\mathbf{k}}$ in $\mathbf{k}$-space.  There are diagonal discontinuity cuts in the $\mathbf{k}$-space.   (d) The improved distribution in panel (c) after a sign flip in every other strip.  Now the discontinuities are removed.   (e) Illustration of the symmetrized Gram-Schmidt process.    }
\label{WannierBloch}
\end{figure*}

For the $\pi$-flux case, the unit cell is doubled with respect to the 0-flux case and hence includes 6 sites (labeled as A, B, C, D, E, and F), as shown in Fig.~\ref{WannierBloch}(a).  Hence, the lowest flat bands are doubly degenerate, which means that there is an arbitrary choice to decompose the two flat bands, since for each $k$ one can arbitrarily choose two orthogonal Bloch vectors in the 2-dimensional degenerate subspace. A sensible choice of basis has to be physically motivated and has to respect certain symmetries.  

For the sake of convenience, we choose loops along the $\mathbf{e}_3$-direction as our preferred flat band basis (total number $N_\text{site}/3$ equals the number of flat-band degeneracy).  The operators $L^{\dag}_{s,\mathbf{R}}$, which create the two types of loops  within each unit cell, can be represented by the original lattice boson operators as:  
\begin{align}
\nonumber L^{\dag}_{1,\mathbf{R}}=&b^{A \dag}_{\mathbf{R}}-b^{B\dag}_{\mathbf{R}}+b^{C\dag}_{\mathbf{R}}-b^{E\dag}_{\mathbf{R}}-b^{E\dag}_{\mathbf{R}-\mathbf{a}_2}+b^{F\dag}_{\mathbf{R}-\mathbf{a}_1} \\
\nonumber &- b^{F\dag}_{\mathbf{R}-\mathbf{a}_1+\mathbf{a}_2}+b^{B\dag}_{\mathbf{R}+\mathbf{a}_2}-b^{A\dag}_{\mathbf{R}+2\mathbf{a}_2}+b^{C\dag}_{\mathbf{R}+\mathbf{a}_2},   \\
\nonumber L^{\dag}_{2,\mathbf{R}}=&b^{B\dag}_{\mathbf{R}+\mathbf{a}_1}-b^{B\dag}_{\mathbf{R}+\mathbf{a}_1-\mathbf{a}_2}-b^{C\dag}_{\mathbf{R}} - b^{C\dag}_{\mathbf{R}+\mathbf{a}_2} + b^{D\dag}_{\mathbf{R}} - b^{D\dag}_{\mathbf{R}+2\mathbf{a}_2}\\
&+b^{E\dag}_{\mathbf{R}} +b^{E\dag}_{\mathbf{R}+\mathbf{a}_2} -b^{F\dag}_{\mathbf{R}}+b^{F\dag}_{\mathbf{R}+\mathbf{a}_2}.
 \end{align} 
Here, the index $s=1,2$ labels the loops encircling left/right hexagons. The wavefunctions of the  two types of loops are already shown in Fig.~3(a).
Here, the lattice vectors $\mathbf{R} \equiv (m,n)$ labels the enlarged 6-site unit cell and the lattice vectors $\mathbf{a}_1$$\equiv$$(1, 0)$ and $\mathbf{a}_2$$\equiv$$(0, 1)$ translate the cells in the two oblique directions [shown in Fig.~\ref{WannierBloch}(a)].  

One can construct two classes of Bloch states by a translationally-invariant superposition (with a particular wavevector $\mathbf{k}$) of the two types of loop states respectively, i.e.~
\be
L^{\dag}_{s,\mathbf{k}}=\sum_{\mathbf{R}} e^{-i \mathbf{k}\cdot \mathbf{R}} L^{\dag}_{s, \mathbf{R}}.
\ee
Here, left/right label $s$$=$$1,2$ can also be thought as the band index and in this particular case labels the two degenerate flat bands.    The generated state $\displaystyle L^{\dag}_{s,\mathbf{k}}\ket{0}$$=$$\sum_{l} u^l_{s,\mathbf{k}} b^{l \dag}_{\mathbf{k}} \ket{0}$ can be represented by a 6-component Bloch vector $\vec{u}_{s, \mathbf{k}}$, where we have $l$$=$$A, B... , F$ and $\displaystyle b^{l \dag}_{ \mathbf{k}}$$=$$\sum_{\mathbf{R}} e^{-i \mathbf{k}\cdot \mathbf{R}} b^{l \dag}_{\mathbf{R}}.$ 
The two Bloch vectors got from the two chosen loop states are represented as
\begin{align}
\nonumber   \vec{u}_{1,\mathbf{k}}=& (1-e^{-i 2\mathbf{k}\cdot \mathbf{R_2}},  -1+e^{-i \mathbf{k}\cdot \mathbf{R_2}},  1+ e^{-i \mathbf{k}\cdot \mathbf{R_2}} , 0, \\
\nonumber& -1-e^{-i \mathbf{k}\cdot \mathbf{R_2}}, e^{i \mathbf{k}\cdot \mathbf{R_1}}-e^{i \mathbf{k}\cdot (\mathbf{R_1}-\mathbf{R_2})} )^T,    \\
\vec{u}_{2,\mathbf{k}}=&  (0,   e^{-i \mathbf{k}\cdot \mathbf{R_1}} - e^{-i \mathbf{k}\cdot (\mathbf{R_1}- \mathbf{R_2}) },  -1- e^{-i \mathbf{k}\cdot \mathbf{R_2}} , \\
\nonumber &1-e^{-i 2 \mathbf{k}\cdot \mathbf{R_2}} ,  1+e^{-i \mathbf{k}\cdot \mathbf{R_2}}, -1-e^{-i \mathbf{k}\cdot \mathbf{R_2}} )^T .
\end{align}

 However, the Bloch vector $\vec{u}_{s, \mathbf{k}}$ is not yet normalized.   We call the normalized Bloch vectors $\vec{\beta}_{s, \mathbf{k}}$, and define the normalized Bloch state as $\displaystyle \ket{\beta_{s, \mathbf{k}}}$$=$$\tilde{L}^{\dag}_{s,\mathbf{k}}\ket{0}$$= $$ \sum_l \beta^l_{s,\mathbf{k}} b^{l \dag}_{\mathbf{k}} \ket{0}$, where the redefined operator $\tilde{L}^{\dag}_{s,\mathbf{k}}$ now becomes canonical bosonic operators satisfying the commutation relation $[\tilde{L}_{s,\mathbf{k}},  \tilde{L}^{\dag}_{s,\mathbf{k'}}]= \delta_{\mathbf{k}, \mathbf{k'}} $.    
Thus, we get a set of orthonormal Bloch states for each of the two flat bands, and can be transformed into two sets of Wannier states as: 
\be
\text{w}^{\dag}_{s, \mathbf{R}} = \frac{1}{\sqrt{N_\text{site}}} \sum_{\mathbf{k}} e^{i \mathbf{k}\cdot \mathbf{R}} \tilde{L}^{\dag}_{s,\mathbf{k}} \equiv \sum_{\mathbf{R'}, l}  w^{l}_{s, \mathbf{R}} (\mathbf{R'}) b^{l \dag}_{\mathbf{R}'} ,
\ee       
where the Wannier wavefunction is given by
\be\label{transformfunction}
w^{l}_{s, \mathbf{R}} (\mathbf{R'}) = \frac{1}{\sqrt{N_\text{site}}} \sum_{\mathbf{k}} e^{i \mathbf{k} \cdot \mathbf{R}} \beta^{l}_{s,\mathbf{k}} e^{-i \mathbf{k} \cdot \mathbf{R}'}.
\ee
Here, the Wannier wavefunction sits on the coordinate $(\mathbf{R'}, l)$.  The coordinate $({\mathbf{R}, s})$ labels where the center of the wavefunction locates.  We note that the more detailed notations,   $\text{w}^{\dag}_{s, \mathbf{R}} $ and $w^{l}_{s, \mathbf{R}} (\mathbf{R'}) $, which we use here, are equivalent to the more compact notations we have used in the main text, namely, $\text{w}^{\dag}_j$ and $w_j (\mathbf{r}) $.  The direct correspondence is $\mathbf{r} \equiv (\mathbf{R'}, l)$ and $j \equiv ({\mathbf{R}, s})$. The two sets of wavefunctions are illustrated in Fig.~\ref{WannierBloch}(b), where the $s$$=$$1$ one encircles only the left hexagons in every unit cell and the $s$$=$$2$ one encircles only the right hexagons. The major part of the real-valued Wannier functions are essentially the two dimer  loop states, which we start with.   The amplitude tail spreads out and decays exponentially along the major axis of the loop which ensures orthogonalization.       
However, the two sets of Wannier functions are not mutually orthogonal to each other (for example, those neighboring ones will still have finite overlap) since the two sets of Bloch vectors are not mutually orthogonalized yet.

For each $\mathbf{k}$, one can orthogonalize the two Bloch vectors through the Gram-Schmidt process, i.e.,
\be
\ket{\gamma_{1,\mathbf{k}}} = \frac{ \ket{\beta_{2,\mathbf{k}}} - {\ket{\mathbf{\beta_{1,\mathbf{k}}}}}  \bket{\beta_{1,\mathbf{k}}}{\beta_{2,\mathbf{k}}}}{\abs{\ket{\beta_{2,\mathbf{k}}} - {\ket{\mathbf{\beta_{1,\mathbf{k}}}}}  \bket{\beta_{1,\mathbf{k}}}{\beta_{2,\mathbf{k}}}}},
\ee
which generates a normalized $\ket{\gamma_{1,\mathbf{k}}}$ orthogonal to $\ket{\mathbf{\beta_{1,\mathbf{k}}}}$.  Similarly, one can get a normalized $\ket{\gamma_{2,\mathbf{k}}}$, which is orthogonal to $\ket{\mathbf{\beta_{2,\mathbf{k}}}}$.   Therefore, one can choose either orthogonal pair as the flat-band basis.  However, from either choice,  the acquired Wannier wavefunctions belonging to the two bands have completely different shapes and hence lose translational symmetry.  To preserve translational symmetry and being closer to a dimer shape, we can make a symmetric superposition as:
\begin{align}\label{symmetrize}
\nonumber\ket{\alpha_{1,\mathbf{k}}} &=\frac{1}{\sqrt{2}} \left[ \ket{\beta_{1,\mathbf{k}}} +e^{i \theta_{\mathbf{k}}} \ket{\gamma_{2, \mathbf{k}}} \right],  \\
 \ket{\alpha_{2,\mathbf{k}}} &=\frac{1}{\sqrt{2}} \left[ \ket{\beta_{2,\mathbf{k}}} +e^{i \theta_{\mathbf{k}}} \ket{\gamma_{1, \mathbf{k}}} \right],
\end{align}
where $\ket{\alpha_{1,\mathbf{k}}}$ and $\ket{\alpha_{2,\mathbf{k}}}$ are the mutually-orthogonal sets of Bloch states we choose.    The additional free choice of phase factor $e^{i \theta_{\mathbf{k}}}$ will give rise to different Wannier states. The sensible choice of the phase factor makes sure that the Bloch vectors are analytically continuous in k-space, which ensures that the generated Wannier function is exponentially localized \cite{Kohn:1973wc} and hence is more compact. 

The Bloch states generated from the Gram-Schmidt process,  $\ket{\gamma_{s,\mathbf{k}}}$, are unfortunately not analytically continuous.   For example, as shown in Fig.~\ref{WannierBloch}(c), the $A$-component of one of the Bloch vectors, $\gamma^A_{1,\mathbf{k}}$, has a diagonal discontinuity cut in its real part.   Same cut happens for its imaginary part and most of the other components of $\ket{\gamma_{1,\mathbf{k}}}$, and $\ket{\gamma_{2,\mathbf{k}}}$.

Thus, one has to take advantage of the additional phase factor $e^{i \theta_\mathbf{k}}$ to remove the discontinuity.  In this particular case, a simple sign flip of every other strip in the $k$-space, which can be expressed as a square-wave function: $\text{sgn}[\sin[(k_x, k_y)\cdot(1, \sqrt{3})]]$, is able to remove the discontinuity [see Fig.~\ref{WannierBloch}(d)].   In addition,  we employ an extra phase factor $e^{i \mathbf{k} \cdot \mathbf{a_2}}$ to ensure no breaking of TR symmetry and closeness in shape to the dimer loop state.    Thus, our choice of phase factor $e^{i \theta_\mathbf{k}}$ for Eq.~(\ref{symmetrize}) is  $e^{i \mathbf{k} \cdot \mathbf{a_2}}\text{sgn}[\sin[(k_x, k_y)\cdot(1, \sqrt{3})]]$ (a $\mathbf{k}$-independent relative sign or phase factor does not affect the probability distribution of the Wannier functions).   This particular choice yields the complete Wannier basis illustrated in Fig.~3(b). 

Our Wannier orbitals (WOs) preserve the mirror symmetry (in terms of probability) with respect to its major axis, similar to the original loop orbitals (LOs) which they are based on.  However, due to the additional phase factor we choose to keep the analytical continuity, the mirror symmetry along the minor axis is slightly broken.  We can see that the lower part of the WO has slightly larger probability than the upper part.   If we replace part of the phase factor $e^{i \mathbf{k} \cdot \mathbf{a_2}}$  with  $e^{-i \mathbf{k} \cdot \mathbf{a_2}}$, the shape of the WO will be flipped with respect to the minor axis, namely the higher part will have larger probability.     We also note that we do not claim that we have found the maximally compact WOs, even though the construction is based on the maximally compact LOs.      In general, it should be possible to numerically/analytically determine such maximally compact WO, which also preserve both types of mirror symmetries.    Thus, our current approach is just a simple mathematical construction, which aims to approximate the maximally compact WOs,  since the shape we have acquired is not too far from the original LOs, which they are based on.    

Finally, we note that, since we have successfully found a complete orthogonal Bloch or Wannier basis from superposition of the dimer loop states,  we have explicitly shown the completeness of the loop states which is mentioned in property 2 of Sec.~\ref{lle}.

\section{Summary of terms in the effective Hamiltonian}\label{summaryeffective}

\begin{figure*}
\includegraphics[width=2\columnwidth]{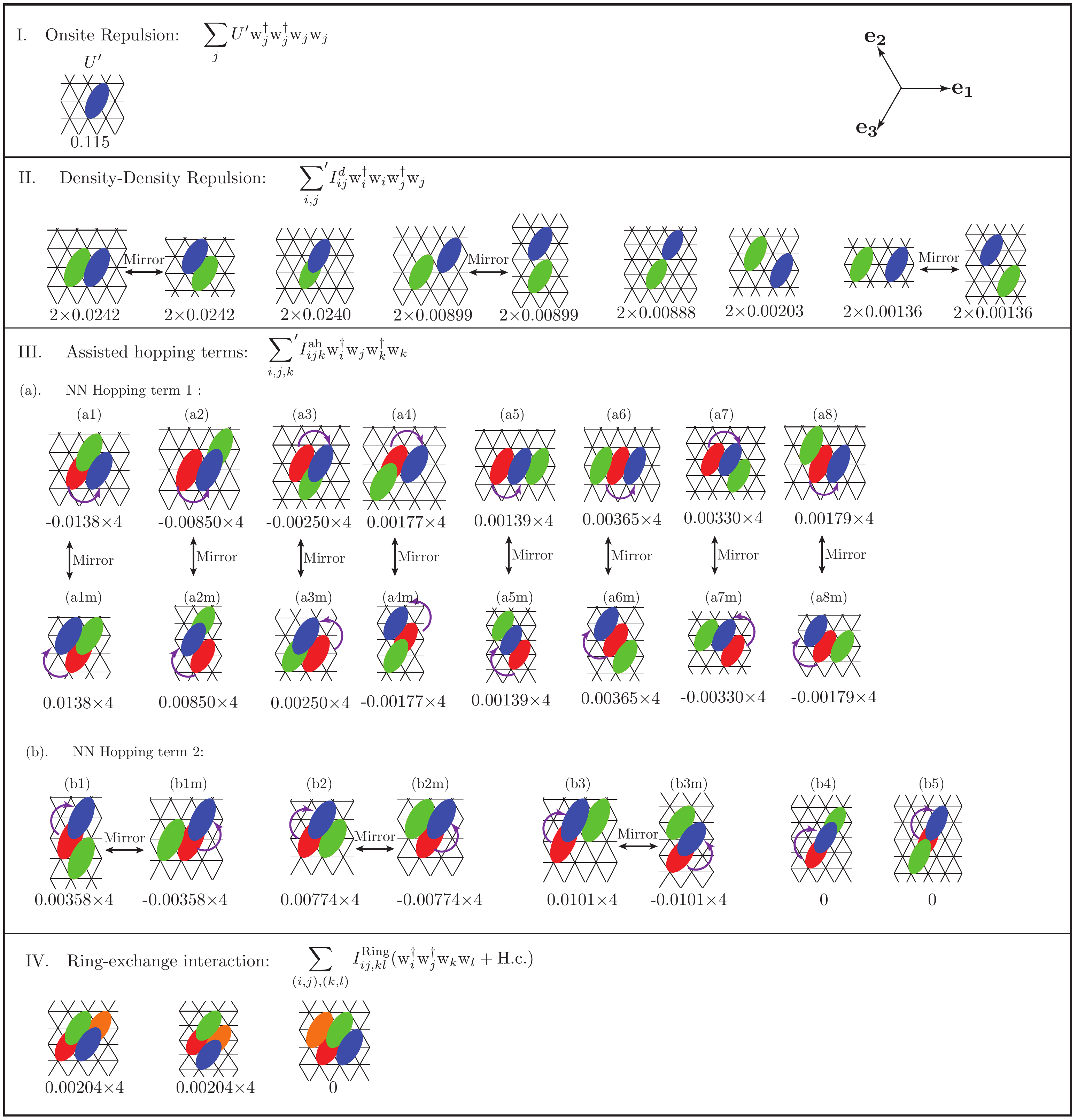}
\caption{(color online).  Table summarizing the leading terms and their coefficients (in units of Hubbard interaction strength $U$) of the effective onsite repulsion and the other three types of effective interactions which survive under the hard-core constraint. }
\label{WannierTable}
\end{figure*}

Here we classify all types of effective interaction  $\sum_{i j k l} I_{ i j k l} \text{w}^\dagger_i \text{w}^\dagger_j \text{w}_k \text{w}_l$, not limited by the hard-core constraint. 
The types of terms are following:

(1) On-site repulsion:
\be
V_\text{onsite}=\sum_j U'  \text{w}^\dagger_j \text{w}^\dagger_j \text{w}_j \text{w}_j,
\ee
where $U'=I_{jjjj} \approx 0.11U$ is the largest energy scale in the effective Hamiltonian.

(2) Density-density repulsion:
\be
V_\text{DD}=\sum_{(i,j)} 2 I^d_{ij}  \text{w}^\dagger_i \text{w}_i \text{w}^\dagger_j \text{w}_j={\sum_{i,j}}' I^d_{ij}  \text{w}^\dagger_i \text{w}_i \text{w}^\dagger_j \text{w}_j,
\ee
where $(i,j)$ means sum over pairs of sites ($i \neq j$). Thus, $(i,j)$ and $(j,i)$ correspond to the same term and should not be double counted.  Now we determine the coefficients $I^d_{ij}$ of the effective interaction.   Four terms in the effective interaction correspond to the pair $(i,j)$, namely 
$I_{ijji} \text{w}^\dagger_i \text{w}^\dagger_j \text{w}_j \text{w}_i+ I_{ijij} \text{w}^\dagger_i \text{w}^\dagger_j \text{w}_i \text{w}_j+ I_{jiji} \text{w}^\dagger_j \text{w}^\dagger_i \text{w}_j \text{w}_i+ I_{jiij} \text{w}^\dagger_j \text{w}^\dagger_i \text{w}_i \text{w}_j = 4 I_{ijji} \text{w}^\dagger_i \text{w}_i \text{w}^\dagger_j \text{w}_j$.  Thus, we get $I^d_{ij}$$=$$ 2 I_{ijji}$.

(3a) Onsite pair-hopping (involving two different sites):
\be
V_\text{PHa}=\sum_{(i,j)} I^p_{ij} ( \text{w}^\dagger_i \text{w}^\dagger_i \text{w}_j \text{w}_j + \text{H.c.})
= {\sum_{i,j}}' I^p_{ij} \text{w}^\dagger_i \text{w}^\dagger_i \text{w}_j \text{w}_j  , 
\ee
where $I^p_{ij} = I_{iijj}$. 

(3b) Off-site pair-hopping (involving three different sites):
\begin{align}
\nonumber V_\text{PHb}=&\sum_{(i,j)}\sum_{k\neq i,j} [ I^p_{ijk} ( \text{w}^\dagger_i \text{w}^\dagger_j +  \text{w}^\dagger_j \text{w}^\dagger_i ) \text{w}_k \text{w}_k + \text{H.c.}]   \\
= &{\sum_{i,j,k}}'  I^p_{ijk} [ \text{w}^\dagger_i \text{w}^\dagger_j  \text{w}_k \text{w}_k + \text{H.c.}],
\end{align}
where $I^p_{ijk} =  I_{ijkk}$. 

(4a) Assisted-hopping (involving three different sites):
\begin{align}
V_\text{AHa}=&\sum_{(i,j)} \sum_{k\neq i,j} I^{\text{ah}}_{ijk} ( \text{w}^\dagger_i \text{w}_j \text{w}^\dagger_k \text{w}_k + \text{H.c.}) \\
=&{\sum_{i,j,k}}' I^{\text{ah}}_{ijk}  \text{w}^\dagger_i \text{w}_j \text{w}^\dagger_k \text{w}_k. 
\end{align}   
Four terms (and their H.c.) in the effective interaction correspond to the pair $(i,j)$, namely 
$I_{ikjk} \text{w}^\dagger_i \text{w}^\dagger_k \text{w}_j \text{w}_k+ I_{ikkj} \text{w}^\dagger_i \text{w}^\dagger_k \text{w}_k \text{w}_j+ I_{kijk} \text{w}^\dagger_k \text{w}^\dagger_i \text{w}_j \text{w}_k+ I_{kikj} \text{w}^\dagger_k \text{w}^\dagger_i \text{w}_k \text{w}_j = 4 I_{ikjk} \text{w}^\dagger_i \text{w}_j \text{w}^\dagger_k \text{w}_k$.  Thus, we get $I^{\text{ah}}_{ijk} = 4 I_{ikjk}$ ($i \neq j \neq k$).

(4b) Assisted-hopping (involving only two different sites):
\begin{align}
\nonumber V_\text{AHb}=&\sum_{(i,j)} [ I^{\text{ah}}_{ijj} ( \text{w}^\dagger_i \text{w}_j + \text{H.c.} )\text{w}^\dagger_j \text{w}_j +  I^{\text{ah}}_{jii} (\text{w}^\dagger_i \text{w}_j + \text{H.c.}) \text{w}^\dagger_i \text{w}_i] \\
=&{\sum_{i,j}}' I^{\text{ah}}_{ijj}  (\text{w}^\dagger_i \text{w}_j + \text{H.c.}) \text{w}^\dagger_j \text{w}_j .
\end{align}
Here, we have $I^{\text{ah}}_{ijj}= 2 I_{ijjj}$, due to the presence of two terms of each type, e.g.  $ I_{ijjj} \text{w}^\dagger_i \text{w}_j \text{w}^\dagger_j \text{w}_j+  I_{jjij} \text{w}^\dagger_j \text{w}_j \text{w}^\dagger_i \text{w}_j= 2 I_{ijjj} \text{w}^\dagger_i \text{w}_j \text{w}^\dagger_j \text{w}_j$.

(5) Ring-exchange interaction:
\be
V_\text{Ring}=\sum_{(i,j),(k,l)} I^\text{Ring}_{ij,kl} ( \text{w}^\dagger_i \text{w}^\dagger_j \text{w}_k \text{w}_l + \text{H.c.}). 
\ee
Here, $\sum_{(i,j),(k,l)}$ means each term in the sum selects two pair of sites (none of the four sites coincide), one pair with creation operators and the other pair with annihilation operators.  Due to the fact that $(i,j)$ and $(j,i)$ correspond to the same thing, there are $2 \times 2=4$ terms from effective interaction correspond to the same type.  Thus, we get $I^\text{Ring}_{ij,kl} = 4 I_{ijkl}$.   Note that, for each plaquette $(i,j,k,l)$, there are $(^4_2)=4$ different types of terms, due to the different choices of creation and annihilation operators. 

After imposing the hard-core approximation, namely replacing the Wannier operators $\text{w}^\dag_j$ by the Pauli operator $\sigma^+_j$, only the density-density repulsion (2),   assisted hopping involving three different sites (4a) and the ring-exchange interaction (5) survive.   Other types of terms vanish due to the doubling of Pauli operators on the same site, namely $(\sigma^+_j)^2=0$.  We list the leading terms of these  types in Fig.~\ref{WannierTable}.

There are certain pairs of effective terms illustrated in Fig.~\ref{WannierTable} which are associated with a mirror reflection along the major axis ($\mathbf{e}_3$).  For the density-density repulsion, terms in such a pair have exactly the same coefficients.   For the assisted hopping, terms in the mirror pair have the same magnitude for coefficients, but may have opposite signs.   In particular, for the assisted hopping in row (b) (hopping along the major axis $\mathbf{e}_3$), terms in all the mirror pairs have opposite signs.   One can see that for uniform density case (e.g.~nematic superfluid) and within the mean-field approximation, the assistive number operator of the dimer (green) can be replaced by a constant number.   Thus, the mirror terms will cancel out exactly and there will be no effective hopping along the major axis ($\mathbf{e}_3$) of the dimer.   The finite effective hopping of the dimers only occur in the other two directions ($\mathbf{e}_1$ and $\mathbf{e}_2$).    Therefore, in a nematic superfluid, the hopping of the bosons is anisotropic.

\section{Details of the self-consistent mean-field theory on a large lattice}\label{algorithm}

With the decoupling of effective interaction mentioned in the main text, we search for the self-consistent solution on a large periodic lattice (torus) with randomized initial distribution of the mean-field order parameters.  We use a local update algorithm: 
\begin{enumerate}
\item
randomly pick up a site $i$ in each step and find the local ground state $\ket{g}_i$ of the decoupled Hamiltonian $h_i (\{\psi_j \}, \{n_j\} )$;  
\item
Calculate the expectation values of the corresponding Wannier operator and Wannier number operator and use them as updated order parameters for site~$i$, i.e.~$\psi_i=_i$$\bra{g} \text{w}_i \ket{g}_i$ and   $n_i=_i$$\bra{g} \text{n}_i \ket{g}_i$;  
\item
Repeat the previous two steps until the order parameters on each site have converged and hence reach the self-consistent mean-field solution.
\end{enumerate}

 \begin{figure*}
\includegraphics[width=1.4 \columnwidth]{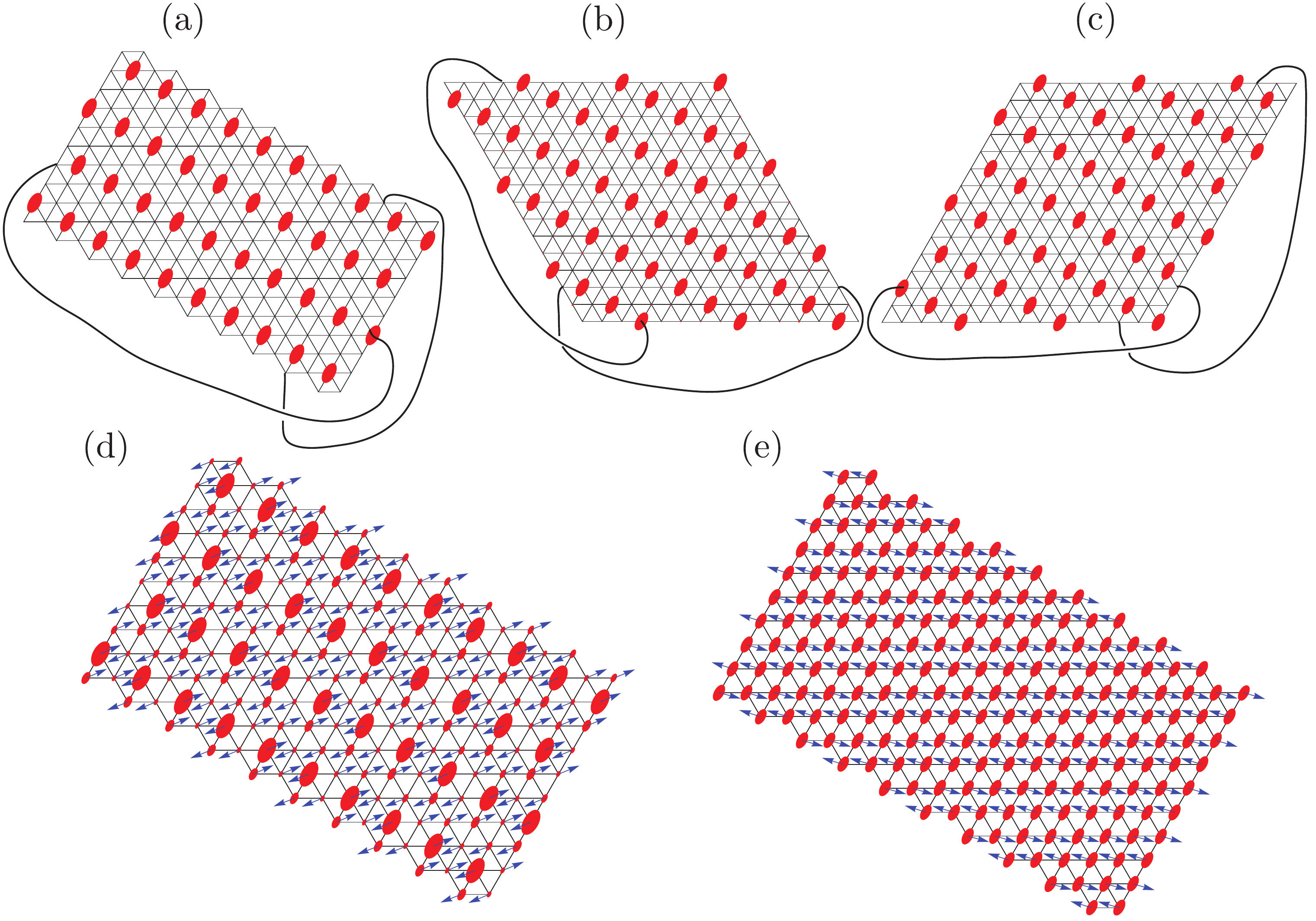}
\caption{(color online).   (a,b,c) The nematic Wigner crystal ground states from self-consistent mean-field calculations on three different lattices. The periodic boundary condition is illustrated by the wires. The lattices contain 200, 225 and 225 sites respectively. (d,e) The nematic supersolid and nematic superfluid states calculated form the self-consistent mean-field theory on the periodic lattice shown in panel (a).}
\label{self-consistency}
\end{figure*}

We have done calculations on lattices with different types of geometry, as shown in Fig.~\ref{self-consistency}(a-c) (with the periodic boundary conditions illustrated by the wires).  For all the cases, the nematic Wigner crystal states are produced (as shown in the figure),  as well as the other two phases.   However, the convergence time of the simulation differs from the lattice geometry.   We find lattice (a) has the fastest convergence, possibly due to the fact that it respects the mirror symmetry of the dimers along their major axis.    Therefore, we use lattice (a) to calculate all the curves shown in Fig.~4(a).  For each chemical potential,  we do several independent simulations with different randomized initial conditions and pick the one with the lowest energy as our solution.   This is due to the fact that sometimes the configuration may be trapped in certain local energy minimum and stops evolving into the true mean-field solution.   The complete image of the simulation results (a nematic supersolid state and a nematic superfluid state) from Fig.~4(c) and 4(d) is shown in Fig.~\ref{self-consistency}(d) and ~\ref{self-consistency}(e).  

\vspace{0.5in}

\section{Calculation of momentum distribution for nematic superfluid and supersolid phases}\label{TOFcalculation}

For the nematic superfluid and supersolid phases, the calculation in the above manner becomes more sophisticated since the correlation between particles on different loops also contributes significantly.   A more convenient way is to do the calculation in the Wannier basis, since the mean-field wavefunction for these two types of ground states is already known and can be expressed by Eq.~\eqref{wavefunction}.
The coefficients $f_{j,0}$ and $f_{j,1}$ are determined by the numerical self-consistent solution discussed in Sec.~\ref{MFT} and Appendix \ref{algorithm}.    Now we see that the single-particle density matrix takes the following form,
\begin{widetext}
\be\label{correlation1}
\boket{\psi}{b^\dag_{\mathbf{r}_1} b_{\mathbf{r}_2} }{\psi} =  \boket{0}{  \prod_{j_1}(f^*_{j_1, 0}+f^*_{j_1, 1} \text{w}_{j_1})    \sum_{j,j'} [ w^*_j(\mathbf{r}_1) \text{w}^\dag_{j} ][ w_{j'}(\mathbf{r}_2) \text{w}_{j'}  ] \prod_{j_2} (f_{j_2, 0}+f_{j_2, 1} \text{w}^\dag_{j_2}) }{0}.
\ee
\end{widetext}
Here, we have already rewriten the original bosonic operators in terms of Wannier operators by flat-band projection Eq.~\eqref{proj}.  One can find that the nonzero contributions in Eq.~(\ref{correlation1}) and simplify it to:
\begin{widetext}
\begin{align}
\nonumber  & \boket{\psi}{b^\dag_{\mathbf{r}_1} b_{\mathbf{r}_2} }{\psi}   =  \sum_{j}    \abs{f_{j, 1}}^2   (1- \abs{f_{j, 0}}^2)  \bra{0} \mathbf{w}_j b^\dag_{\mathbf{r}_1}  b_{\mathbf{r}_2} \mathbf{w}^\dag_j \ket{0}+ {\sum_{j, j'}}'  f_{j, 0}  f^*_{j, 1}   f^*_{j', 0}  f_{j', 1}  w^*_{j}(\mathbf{r}_1)  w_{j'}(\mathbf{r}_2).
\end{align} 
\end{widetext}

The first term originates from the correlation between particles in the same Wannier orbital.   It is proportional (up to a constant prefactor) to the single-particle density matrix we calculated in Sec.~\ref{momTOF} for the NWC states [inside Eq.~(\ref{NWCcorrelation})] with the approximation that replaces the loop orbitals ($L_j$) with the Wannier orbitals ($\mathbf{w}_j$).   For either nematic superfluid or nematic supersolid states, the background contribution to the momentum distribution from this term (after a Fourier transform), denoted by $\langle n^{(1)}_\mathbf{q} \rangle$, does not change up to a constant prefactor.  More concretely, this background contribution for superfluid and supersolid states is reduced since a certain portion of the occupation moves to the condensation peaks as will be explained in the following paragraphs.

The second term originates from the correlation between particles in different Wannier orbitals.  Clearly, it only becomes nonzero when $f_{j,0}, f_{j,1}, f_{j',0}$ and $f_{j',1}$ are all nonzero.
Thus, this term disappears for the NWC phase where each Wannier orbital is in a number state (Fock state), either with $n=0$ or $n=1$.  It becomes nonzero when certain Wannier orbitals are in superposition states, which implies superfluidity and presence of off-diagonal long-range order.   The Fourier transform of the second term can be expressed as
\begin{widetext}
\be
\langle  n^{(2)}_\mathbf{q} \rangle = \frac{1}{N_\text{site}}  {\sum_{j,j'}}'     f_{j, 0}  f^*_{j, 1}   f^*_{j', 0}  f_{j', 1}  e^{i \mathbf{q} \cdot (\mathbf{R}_j - \mathbf{R}_{j'})}   \sum_{\mathbf{r}_1} e^{i \mathbf{q} \cdot (\mathbf{r}_1 - \mathbf{R}_j)} w_j(\mathbf{r}_1)    \sum_{\mathbf{r}_2} e^{- i \mathbf{q} \cdot (\mathbf{r}_2 - \mathbf{R}_{j'})} w^*_{j'}(\mathbf{r}_2).           
\ee
\end{widetext}
Here, $\mathbf{R}_j$ refers to the coordinate of the center of Wannier orbital $j$ [equivalent to the composite label ($\mathbf{R}, s$) in Appendix~\ref{construction}].  Note that the Fourier transform of the Wannier function is the Bloch function [inverse transform of Eq.~\eqref{transformfunction}], therefore we get 
\be
 \frac{1}{\sqrt{N_\text{site}}}  \sum_{\mathbf{r}_1} e^{i \mathbf{q} \cdot (\mathbf{r}_1 - \mathbf{R}_j)} w_j(\mathbf{r}_1)  =  \sum_l  \tilde{\beta}^l_{s(j)} (\mathbf{q}).
\ee
Here we have already performed an additional summation over the ten Bloch vector components labeled by $l$ (note that $\sum_{\mathbf{r}_1}$ actually corresponds to a double sum $\sum_{\mathbf{R}_1} \sum_l$, where $\mathbf{R_1}$ labels all the unit cells and $l$ labels the sites in each unit cell). The label $s(j)$ (which depends on the Wannier center $j$) is the band index.  In the $\pi$-flux case, it labels left ($s=1$) and right ($s=2$) Bloch/Wannier states (two degenerate flat bands) as mentioned in Appendix~\ref{construction}.  Therefore, we can further simplify the above expression as
\begin{widetext} 
\begin{align}
 \langle  n^{(2)}_\mathbf{q} \rangle  
= & N_\text{site} \left[ \frac{1}{\sqrt{N_\text{site}}} \sum_j   \psi^*_j  e^{i \mathbf{q} \cdot \mathbf{R}_j }      \sum_l  \tilde{\beta}^l_{s(j)} (\mathbf{q})    \right]   \left[  \frac{1}{\sqrt{N_\text{site}}} \sum_{j'}  \psi_{j'} e^{-i \mathbf{q} \cdot \mathbf{R}_{j'} }     \sum_l  \tilde{\beta}^{*l}_{s(j')} (\mathbf{q})         \right].
\end{align}
\end{widetext} 
Here, we have expressed the result in terms of $\psi_j$$=$$\langle \mathbf{w}_j  \rangle$$=$$e^{i \varphi_j}\frac{1}{2} \sin(\theta_j)  $, namely the mean-field superfluid order parameter parameterized by the two Bloch-sphere angles of a pseudo spin-$\frac{1}{2}$ (equivalent to hard-core boson).   We can also express the mean-field parameters as  $f_{j, 0}$$=$$\cos (\theta_j/2)$ and $f_{j, 1}$$=$$e^{i \varphi_j} \sin (\theta_j/2)$,  which implies the relation $\psi_j$$=$$ f_{j, 0} f_{j, 1}$. 

Now we first consider the simple 0-flux case, when there is only one flat band.  Hence, the summation of the Bloch components is site-independent.   We can factorize the above expression into
\begin{align}
\langle  n^{(2)}_\mathbf{q} \rangle_{\text{0-flux}}  = N_\text{site} \big|\psi (\mathbf{q})\big|^2 \cdot  \bigg|\sum_{l=1}^3  \tilde{\beta}^l (\mathbf{q}) \bigg|^2,
\end{align}
where $\psi (\mathbf{q})$ is the Fourier transform of the superfluid order parameter $\psi_j$ on the effective triangular lattice.   Thus, for the superfluid and supersolid phases, the momentum distribution will have the same set of delta-function peaks as those acquired from Fourier transforming the superfluid order parameter on the effective triangular lattice.  However, an additional form factor $ \abs{\sum_l  \tilde{\beta}^l (\mathbf{q}) }^2$ changes the weights of the delta-function peaks (e.g. in the supersolid phases).   The simplicity here is essentially due to the fact that, for the 0-flux kagome lattice and the effective triangular lattice have the same reciprocal lattice.

In the $\pi$-flux case, there are two degenerate bands due to doubling of unit cell size, and the size of the 1st-Brillouin zone is reduced by half.  In this case we can split the effective triangular lattice into two sub-lattices:  one corresponds to the centers of the left Wannier states, labeled by $j_L$, and the other corresponds to the centers of the right Wannier states, labeled by $j_R$.   Note that these two sub-lattices correspond to the same reciprocal lattice of the $\pi$-flux kagome lattice, the size of which is reduced by half.  Thus, we can re-express the momentum distribution as:
\begin{align}
\nonumber \langle  n^{(2)}_\mathbf{q} \rangle_\text{$\pi$-flux}
=&    N_\text{site} \bigg|  \psi_L (\mathbf{q})   \sum_{l=1}^6  \tilde{\beta}^{*l}_{s=1} (\mathbf{q}) + \psi_R (\mathbf{q})    \sum_{l=1}^6  \tilde{\beta}^{*l}_{s=2} (\mathbf{q}) \bigg|^2  ,
\end{align}
where $\psi_L (\mathbf{q})$ and $\psi_R (\mathbf{q})$ corresponds to the Fourier transform of the order parameters on the left and right effective sub-lattices.   The delta-function peaks of the momentum distribution come from the momentum peaks of the superfluid order parameter of both sub-lattices.

The continuous backgrounds in the three phases (nematic Wigner crystal, nematic superfluid and nematic supersolid) have the same pattern up to a constant prefactor.    On top of that, the nematic superfluid and supersolid have additional delta-function peaks in their momentum distributions, originating from $ \langle  n^{(2)}_\mathbf{q} \rangle$.  While the superfluid phase has only one peak in each Brillouin zone, the supersolid phase actually has five peaks in 
each Brillouin zone.

\section{Localized orbitals preserving $C_6$ symmetry}\label{energetics}

In Sec.~\ref{nematic} we have constructed a Wannier basis close to the maximally compact dimer LOs. Our effective flat-band projected Hamiltonian is represented on this basis.   In principle, there are an infinite number of Wannier bases one can choose to construct effective Hamiltonians.   These effective Hamiltonians are equivalent and only differ by a basis change.   However, the mean-field ansatz and the mean-field decoupled Hamiltonian [see Eq.~\eqref{decoupling}] constructed from different bases are not equivalent.      Therefore, the nematic phases acquired in Sec.~\ref{nematic} is a direct consequence of choosing a mean-field ansatz based on anisotropic WOs oriented in the same direction.

\begin{figure*}
\includegraphics[width=1.4\columnwidth]{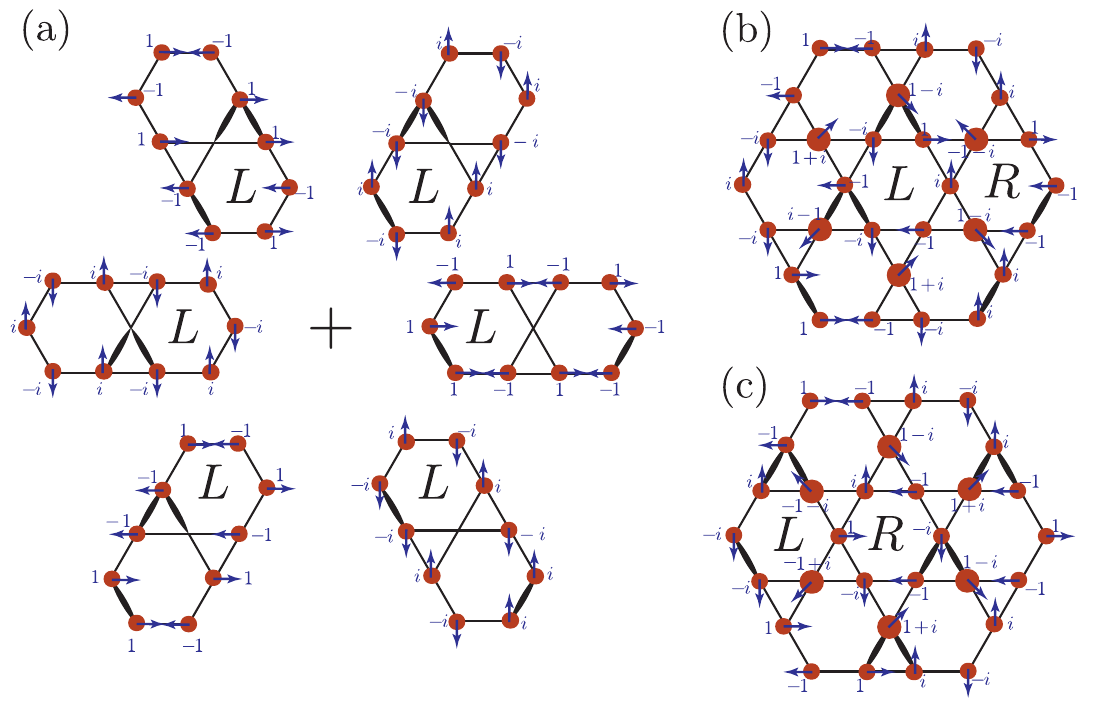}
\caption{(color online).  (a)  Six dimer loop orbitals (with amplitudes labeled for each site) sharing the same left hexagon.   (b) A $C_6$-symmetric localized orbital centered on the left hexagon. The orbital is superimposed by the six dimer loop orbitals in panel (a).  (c) A $C_6$ symmetric localized orbital centered on the right hexagon. The orbital has the same probability distribution as the one in panel (b).  }
\label{symmetric}
\end{figure*}

Here we show that it is actually also possible to construct a localized orbital which preserves the $C_6$ rotational symmetry of probability distribution.  The most compact $C_6$-symmetric orbital can be constructed by the superposition of six dimer loop orbitals as shown in Fig.~\ref{symmetric}.   In panel (a), we have six dimer loop orbitals (with the amplitude labeled for each site) sharing the same left hexagon.   The superposition of the six orbitals generate a $C_6$-symmetric orbital in panel (b), centered on the left hexagon.   We note that in order to preserve the $C_6$-symmetry of probability, one has to introduce relative phase between the orbitals, and in this example a $\pi/2$ phase. This leads to the imaginary number $i$ in certain amplitudes and breaking of TR symmetry.  In this case, the time-reversal partner is obtained by turning $i$ into $-i$.  This orbital also preserves the $C_6$-symmetry and has the opposite current.
The reason of TR-symmetry breaking is that the $C_6$-symmetric orbital includes seven (odd number) $\pi$-fluxes, which gives rise to the doubly degenerate eigenstates with opposite chirality.   A similar construction can generate a $C_6$-symmetric orbital centered on the right hexagon, by the superposition of the six surrounding dimer loops.   Thus, for each unit cell, one can generate a $C_6$-symmetric orbital centered on right hexagons with the same probability distribution.   These orbitals form a complete and spatially uniform basis of the gapped flat band and can also be orthogonalized to form Wannier orbitals.

The physical reasoning on the choice of maximally compact dimer LOs/WOs, instead of $C_6$-symmetric ones, is based on the energetic consideration that particles can avoid each other by occupying the maximally compact orbitals with at most one particle.  In the case of Wigner crystal and supersolid phases as shown in Fig.~\ref{phasediagram}, the orbitals with large occupation probability do not overlap with each other (ignoring the exponential tails).   Therefore, the energetic advantage of being nematic is obvious.  For the uniform nematic superfluid phase,  all the LOs/WOs are equally populated.  Due to the fact that dimer LOs enclose two hexagons (while the number of dimer LOs are equal to the number of hexagons), there is a significant overlap between neighboring dimer LOs.   Still, the overlap is not as large as the case of more extended $C_6$-symmetric orbitals, as shown in Fig.~\ref{symmetric}.   Therefore, nematicity is still energetically favorable even for the uniform phase.

\end{appendix}

%\bibliography{mybib}

\end{document}